\documentclass[]{aastex62}

\usepackage{graphicx}
\usepackage{amsmath,amssymb,amsfonts,textcomp,array}

\newcommand{\mn}{{Mon.\@ Not.\@ Roy.\@ Ast.\@ Soc.\ }}
\newcommand{\asta}{{Astron.\@ Astrophys.\ }}

\newcommand{\plb}{{Phys.\@ Lett.\@ B\ }}

\newcommand{\etal}{{et al.~}}
\newcommand{\ie}{{i.e.,~}}
\newcommand{\etc}{{e.t.c}}

\newcommand{\eg}{{e.g.,~}} 

\newcommand{\beq}{\begin{equation}}
\newcommand{\eeq}{\end{equation}}
\newcommand{\ber}{\begin{eqnarray}}
\newcommand{\eer}{\end{eqnarray}}
\newcommand{\lleq}{\lower0.9ex\hbox{ $\buildrel < \over \sim$} ~}
\newcommand{\ggeq}{\lower0.9ex\hbox{ $\buildrel > \over \sim$} ~}
\newcommand{\lsim}{\ \lower-1.5pt\vbox{\hhbox{\rlap{$<$}\lower5.3pt\vbox{\hhbox{$\sim$}}}}\ }
\newcommand{\gsim}{\ \lower-1.5pt\vbox{\hhbox{\rlap{$>$}\lower5.3pt\vbox{\hhbox{$\sim$}}}}\ }

\newcommand{\omt}{\Omega_{0 \rm m}}
\newcommand{\omr}{\Omega_{0 \rm r}}

\newcommand{\w}{w_{DE}}
\newcommand{\wde}{w_{DE, {\rm eff}}}
\newcommand{\wdep}{w^{\prime}_{DE, {\rm eff}}}
\newcommand{\csde}{c^2_{sDE,{\rm eff}}}
\newcommand{\wdm}{w_{DM, {\rm eff}}}
\newcommand{\csdm}{c^2_{sDM,{\rm eff}}}
\newcommand{\HH}{{\mathcal H}}
\newcommand{\HHs}{{{\mathcal H}^2}}

\newcommand{\lya}{Ly$\alpha$~}
\newcommand{\lam}{\Lambda}

\submitjournal{ApJ}

\begin{document}

\title{ Are $H_0$ and $\sigma_8$ tensions generic to present cosmological data?}

\correspondingauthor{Ujjaini Alam}
\email{ujjaini.alam@gmail.com}

\author{Archita Bhattacharyya}
\affil{Physics \& Applied Mathematics Unit, Indian Statistical Institute, Kolkata India}

\author{Ujjaini Alam}
\affil{Physics \& Applied Mathematics Unit, Indian Statistical Institute, Kolkata India}
\affil{IRAP, Universit\'e de Toulouse, CNRS, UPS, CNES, Toulouse France}

\author{Kanhaiya Lal Pandey}
\affil{Indian Institute of Astrophysics, Bangalore India}

\author{Subinoy Das}
\affil{Indian Institute of Astrophysics, Bangalore India}

\author{Supratik Pal}
\affil{Physics \& Applied Mathematics Unit, Indian Statistical Institute, Kolkata India}

\begin{abstract}
Yes, for a wide range of cosmological models ($\lam$CDM,
non-interacting $w_z$CDM, $w_z$WDM,  or a class of interacting DMDE). Recently there have been attempts to solve
the tension between direct measurements of $H_0$ and $\sigma_8
\sqrt{\omt}$ from respective low redshift observables and 
indirect measurements of these quantities from the CMB observations. In this work we construct a quasi-model
independent framework that reduces to different classes of
cosmological models under suitable parameters choices.  We test this
parameterization against the latest Planck CMB data combined with 
recent BAO, SNe and direct $H_0$ measurements.  Our analysis reveals that a
strong positive correlation between $H_0$ and $\sigma_8$ is more or
less generic for most of the cosmological model. The present data slightly prefers a phantom equation of
state for DE and a slightly negative effective
equation of state for DM (a direct signature of
interacting models), with a relatively high $H_0$ consistent
with Planck+R16 data and, simultaneously, a consistent 
$\omt$. Thus, even though the tensions cannot be fully resolved,
a class of interacting models with phantom $w_{DE}$ get a slight edge over $w_z$CDM for present data. However, although
they may resolve
the tension between high redshift CMB data and individual low
redshift datasets, these datasets have inconsistencies
between them (\eg between BAO and $H_0$, SNe and BAO, and cluster
counts and $H_0$).
\end{abstract}

	
\section{Introduction}\label{sec:intro}

In the current data driven era of cosmology, one of the major
challenges is to illuminate the dark sector of the universe. Since
visible matter has been found to constitute a tiny fraction of the
total matter content of the universe, we need to comprehend the nature
of dark matter, comprising nearly a third of the total energy
content. Dark energy, the enigmatic negative pressure energy component
that dominates the universe today, and causes its expansion to
accelerate, is an even greater mystery. The standard cosmological
model for the universe is the $\lam$CDM model, where dark matter is
expected to be ``cold'', with an equation of state $w_{DM} = 0$, while
dark energy is represented by the cosmological constant, with a
constant energy density and constant equation of state $\w =
-1$. Current observations are more or less commensurate with this
``concordance'' model \citep{planck15}, with one or two
caveats. However, other models for dark matter and dark energy are yet
to be ruled out. For dark energy especially, constraints on its
equation of state are broad enough that many different models can be
accommodated \citep[see reviews][] {sahni00_de, peeb03_de, paddy03_de,
  sahni04_de, cope06_de, frie08_de, durr08_de, tsuj10_de, noj11_de,
  skor12_de, mort14_de, baha17_de}. Dynamical dark energy models can
be divided into two broad categories. Firstly, one may consider dark
energy as a separate energy component, either a fluid or a scalar
field or multiple scalar fields. In the second approach, the
acceleration of the universe can be explained by introducing new
physics in the gravity sector and modifying Einsteinian gravity. Both
types of models have been studied extensively against observations
\citep{alam10_obs, hols10_obs, laz12_obs, critt12_obs, shaf13_obs,
  clark16_obs, shaf17_obs, melc17_obs, mores17_obs, tink17_obs,
  ratra18_obs, amen18_obs}, and although recent gravity wave
observations have placed tight constraints on a large number of
modified gravity models, many other dark energy models still remain
viable.  Coupled or interacting dark matter-dark energy (DMDE) models
are also in vogue.  Though observations suggest that the dark sectors
are mostly non-interacting, mild interaction between them can not be
ruled out. In these models, a coupling in the dark sector allows
either dark matter particles transfer energy into dark energy, or
conversely, for dark energy to decay into dark matter on the Hubble
time scale. Many different phenomenological forms have been proposed
for the interaction and tested against data \citep{amen99_dmde,
  wands00_dmde, coley00_dmde, hwang02_dmde, comel03_dmde,
  pavon03_dmde, farr04_dmde, amen04_dmde, das06_dmde, bean08_dmde,
  mena10_dmde, wett11_dmde, cope12_dmde, pavan12_dmde, amen12_dmde,
  skor13_dmde, val15_dmde, melc17_dmde, nunes17_dmde, sahni18_dmde},
but it is difficult to discriminate between the different interacting
DMDE models.  Also, for these phenomenological models, the results
crucially depend on the somewhat ad hoc choice of the interaction
term.

The different models for cosmology, be it $\lam$CDM, or models that
fall either under the class of non-interacting $w_z$CDM or interacting
DMDE, or warm dark matter models, are usually constrained against a plethora of observations,
including those of the Cosmic Microwave Background (CMB), Baryonic
Acoustic Oscillations (BAO), Type Ia Supernovae (SNeIa), measurements
of the Hubble parameter $H(z)$ from galaxies, direct measurements of
the Hubble constant $H_0$, and weak and strong lensing.

It is noteworthy that there appear to be some inconsistencies between
different cosmological datasets when analyzed against the concordance
$\lam$CDM model. For example, a major discrepancy between observations
arises in the measured value of the Hubble parameter at present. The
Planck 2015 CMB analysis for the $\lam$CDM 3-neutrino model gives a
value of $H_0=67.3 \pm 1.0$ km/s/Mpc \citep{planck15}. However, the
most recent dataset for direct measurement of $H_0$ \citep{riess16_h0}
obtains a $2.4\%$ determination of the Hubble Constant at $H_0 = 73.24
\pm 1.74$ km/s/Mpc. This value disagrees at around \textbf{$\sim 3
  \sigma$} with that predicted by Planck. This is probably the most
persistent tension between cosmological data sets for $\lam$CDM.
Another major source of tension is in the predicted values of $\omt$
and $\sigma_8$ from CMB and from clusters. From Planck, we obtain the
constraints $\sigma_8 \sqrt{\omt/0.3} = 0.851\pm 0.013$, while the
clusters provide a lower value of $\sigma_8 \sqrt{\omt/0.3} = 0.745\pm
0.039$ \citep{bohr14_s8}, a tension at about $2.5\sigma$. Further,
recent BAO measurements in the \lya forest of BOSS DR11 quasars at
redshift $z=2.34$ \citep{del14_baoh} provide a Hubble parameter of
$H(z=2.34)=222\pm 5$ km/s/Mpc, which is $7 \%$ higher than the
predictions of a flat $\lam$CDM cosmological model with the best-fit
Planck parameters, a discrepancy significant at $2.5 \sigma$. In yet
another departure, the lensing parameter $A_L$ is expected to have the
base value of unity for $\lam$CDM, but has instead constraints of $A_L
= 1.22 \pm 0.1$ from Planck \citep{planck15}. Explanations for these
tension may be found in the errors and systematics in the observations
themselves, \eg different analysis methods used for the low redshift
SNeIa data \citep{efst14_h0, alam17_sn}, possible systematic bias in
scaling relations for clusters \citep{mantz15_s8}, tensions of the
\lya BAO data with lower redshift galaxy BAO data \citep{aub14_bao,
  alam17_mg}, \etc. However, since these tensions seem to exist
largely between the high redshift CMB data and low redshift direct
measurements, this might also be interpreted as a hint to go beyond
the standard $\lam$CDM model and look for new physics which changes
the expansion history either at high redshift (by changing $N_{eff}$,
the radiation content \citep{Karwal:2016vyq}) or at low redshift 
(by changing the dark energy dynamics). In this work we explore if 
a richer dark sector can provide us an alternative explanation for 
these discrepancies.

In order to investigate the above-mentioned issues, we analytically
reconstruct a model-independent approach to address different classes
of cosmological models.  We start with the most general interacting
DMDE scenario that takes into account the maximum number of model
parameters, and construct a framework to deal with the background and
perturbation equations in terms of a set of model parameters (namely,
the equations of state and sound speed for DM and DE).  We also
demonstrate that the concordance $\lam$CDM and non-interacting
$w_z$CDM models turn out to be special cases for this generalized
scenario, with suitable choice of model parameters. Thus, we end up
with a framework which takes into account a wide class of cosmological
models, thereby making our subsequent investigation for the $H_0$ and
$\sigma_8$ discrepancies generic and quasi-model independent. Then we
analyze the current observations against our quasi-model independent
reconstruction of cosmological models followed by a comparison among
different cosmological models ($\lam$CDM, non-interacting $w_z$CDM,
interacting DMDE, warm dark matter models) and the role of each dataset on these class of
models.

The plan of the paper is as follows: In section~\ref{sec:setup}, we
outline the model-independent scheme used to represent different
cosmological models, section~\ref{sec:data} describes the data and
methodology used in the analysis, section~\ref{sec:res} gives the
results, comparison among different models and discussions, and in
section~\ref{sec:concl} we present our conclusions.

\section{General Framework for Different Cosmological Models}\label{sec:setup}

We start with a general theoretical framework where there are two fluids, namely, dark energy and dark matter, which may or may not be interacting with each other, and express a set of working formulae, namely,
the background and perturbation equations, in a general
approach.  As we shall show subsequently, the usual $\lam$CDM,
non-interacting $w_z$CDM, a class of interacting dark sector models as well as warm dark matter models  can be considered subsets of this
generic framework with suitable choice of parameters, thereby making
the analysis a fairly comprehensive framework for a wide class of
different cosmological models.

In this generic setup of (non)interacting dark sectors, different
models of the universe have been suggested in the literature and
tested against data with varying degrees of success. 
There is no clear
theoretical preference for one model over the other, the varied models
naturally come up with different parameter space constraints, and are
therefore difficult to compare. In this work, we aim to recast the evolution equations in a way which allows us to include  a wide class of cosmological models,
namely, $\lam$CDM,
non-interacting $w_z$CDM, a class of interacting dark sector models, as well as warm dark matter models, by
suitably choosing the corresponding parameters.
	
\subsection{Background Equations}
	
The general evolution equations for a two-fluid (DM, DE), interacting cosmological system are obtained from conservation of total energy density to be
\ber
\rho^{\prime}_{DM} + 3 \HH (1+w_{DM})\rho_{DM} &=& -aQ \\
\rho^{\prime}_{DE} + 3 \HH (1+\w)\rho_{DE} &=& aQ \,\,,
\eer
where derivatives are taken with respect to the conformal time, and
$Q$ is the rate of transfer of energy density, \ie the interaction
term. When the interaction term is switched off ($Q=0$), 
we regain the non-interacting DM+DE
scenario, while a non-zero $Q$ implies interaction between DM and DE. Usually, when studying interacting DMDE models, $Q$ is
replaced by some functional form \eg $Q=-\Gamma \rho_{DM}$
\citep{boeh08_dmde}, or $Q=\HH
(\alpha_{DM}\rho_{DM}+\alpha_{DE}\rho_{DE})$ \citep{pav01_dmde}. Many
different interaction terms have been suggested, some motivated
physically, others simple phenomenological parameterizations. On the other hand, $w_{DM}=0$ reduces to standard CDM, while a small non-zero $w_{DM}$ would give us warm dark matter, which may or may not interact with dark energy depending on the value of $Q$.
As we will show subsequently, even though the above two equations represent interacting dark sectors, they have the potential to take into account a wide class
of cosmological models under consideration. 

In order
to encompass both the possibilities of warm dark matter and interacting DMDE with fewer parameters,
as well as to take into account the  usual $\lam$CDM and
non-interacting $w_z$CDM, we recast the above equations to resemble the non-interacting $w_z$CDM 
scenario:
\ber	
\rho^{\prime}_{DM, {\rm eff}} + 3 \HH (1+\wdm)\rho_{DM, {\rm eff}} &=& 0 \\
\rho^{\prime}_{DE, {\rm eff}}+3 \HH (1+\wde)\rho_{DE, {\rm eff}} &=& 0 \,\,,
\eer
with the effective equations of state for dark matter and dark energy
defined by adding the effect of the interaction term $Q$ to the true
dark matter and dark energy equations of state:
\ber 
\wdm &=& w_{DM} + \frac{aQ}{3\HH \rho_{DM}} \\
\wde &=& \w -\frac{aQ}{3\HH \rho_{DE}} \,\,.
\eer
In the interacting scenario, for $Q > 0$, energy is transferred from dark matter to dark energy,
which implies $\wdm > 0$; the effective dark matter redshifts at a
rate faster than $a^{-3}$, and $\wde < \w$; the effective dark energy
has more negative pressure. For $Q < 0$, the opposite happens. In the non-interacting scenario, $\wdm \neq 0$ implies non-cold dark matter.

We note here that, for a constant $\wdm-w_{DM}$, this approach takes care of a class of interacting dark sector models where $Q \propto \HH \rho_{DM}$. Apart from this class of interacting models, this approach also has the added advantage that it boils down to different class of dark sector models by suitable choices of its parameters namely, $\wdm$ and $\wde$:

\begin{itemize}
	\item $\wdm=0, ~~ \wde=-1$ ($\Lambda$CDM),
	\item $\wdm=0, \wde < -1$ (phantom), $>-1$ (non-phantom), (non-interacting $w_z$CDM, depending on scalar field or modified gravity models),
	\item $\wdm \neq 0$, (warm dark matter models or a class of interacting dark sector models).
\end{itemize}

Strictly speaking, although $\wdm$ and $\wde$ are independent parameters for all other cosmological models  ($\Lambda$CDM, non-interacting $w_z$CDM, modified gravity and warm dark matter models), they are not strictly independent free parameters for the interacting DMDE models under consideration, because of the coupling term $Q$. However, one cannot  have any a priori knowledge on the interaction term from theoretical  perspectives alone, 
even if there is any such interaction between dark matter and dark energy. In order to have an idea on the interaction, one needs to take shelter of
observational data. 
As it will be revealed in due course, 
observational data puts stringent constraints on any possible interaction, and DMDE interaction, if any, would be really feeble, 
deviating from $\wdm =0$ by a very tiny amount at the most, so that we could  {\it effectively}  decouple the equations of state.
As a result, this parametrization allows us to consider them as independent parameters for all practical purpose. This is what we are going to consider
in the present article.

\subsection{Linear Perturbations}

In this approach, the perturbation equations need to
be similarly recast in terms of effective equations of state for dark
matter and dark energy, so that the interaction term $Q$  does not
explicitly appear in them (or, in turn, the effects of warm dark matter, if any, becomes obvious). Scalar perturbations on a flat FRW metric
are given by,
\beq
ds^2=a^2\{-(1+2\psi)d\eta^2+2\partial_i B d\eta dx^i+[(1-2\phi)\delta_{ij}+2\partial_i\partial_jE]dx^idx^j\} \,\,. 
\eeq

The energy-momentum tensor for the dark sector is given by
\begin{equation}
T^{\mu}_{\nu}=(\rho +P)u^\mu u_{\nu} +P\delta^{\mu}_{\nu} \,\,,
\end{equation}
where $\rho = \bar{\rho}+\delta\rho$, $P = \bar{P}+\delta P$, the
background 4-velocity is $\bar{u}^{\mu}=a^{-1}\delta^{\mu}_0$ and the
perturbed 4-velocity by $u^{\mu}=a^{-1}(1-\psi,\partial^i v),
u_{\mu}=a(-1-\psi,\partial_i [v+B])$, with $v$ as the peculiar
velocity potential. We adopt the synchronous gauge for which
$\psi=B=0, \phi=\eta$ and $k^2E=-h/2-3\eta$.

For a(n) (un)coupled dark sector scenario, the pressure perturbation for
each component is
\begin{equation}
\delta P_i=c_{ai}^2\delta\rho_i +(c_{si}^2 - c_{ai}^2)[3\mathcal{H}(1+w_{i, {\rm eff}})\rho_i]\frac{\theta_i}{k^2} \,\,,
\end{equation}
where $i=DM, DE$. Therefore the background coupling enters $\delta
P_i$ through the term $w_{i, {\rm eff}}$. The effective sound speed
$c_{si, {\rm eff}}$ of a fluid in its rest frame is then defined as,
\begin{equation}
c^2_{si, {\rm eff}}=\frac{\delta P_i}{\delta\rho_i} \,\,,
\end{equation}
and adiabatic sound speed as,
\begin{equation}
c_{ai, {\rm eff}}^2=\frac{P_i'}{\rho^{\prime}_i}=w_{i, {\rm eff}}+\frac{w^{\prime}_{i, {\rm eff}}}{\rho^{\prime}_i/\rho_i} \,\,.
\end{equation}
It is noteworthy to point out here that that the effective sound
speeds reduce to standard sound speeds of non-interacting $w_z$CDM and
$\lam$CDM as soon as the interaction term is switched off.

Using the above definitions, we may now write down the effective
perturbed evolution equations for DM and DE as,
\ber
&& \delta^{\prime}_{DM} + 3\HH (\csdm - \wdm) \delta_{DM} + (1 + \wdm) \theta_{DM} + 9 \HHs [(1 + \wdm)(\csdm - \wdm)] \frac{\theta_{DM}}{k^2} \nonumber\\
&&=(1 + \wdm) \frac{h^{\prime}}{2} \\
&& \theta^{\prime}_{DM} +\HH (1 - 3\csdm) \theta_{DM} - \frac{\csdm}{(1+\wdm)} k^2 \delta_{DM} = 0\\
&& \delta^{\prime}_{DE}+3\HH (\csde-\wde) \delta_{DE} + (1 + \wde) \theta_{DE} +9 \HHs \left[(1 + \wde)(\csde - \wde) + \frac{\wdep}{3 \HH}\right]\frac{\theta_{DE}}{k^2} \nonumber\\
&&= (1 + \wde) \frac{h^{\prime}}{2} \\
&& \theta^{\prime}_{DE}+ \HH (1 - 3\csde) \theta_{DE} - \frac{\csde}{(1+\wde)} k^2 \delta_{DE} = 0 \label{eq:pert_v}\,\,.
\eer
We note here that, in the synchronous gauge, DM particles are
typically taken as gauge coordinates so that $\theta_{DM}$
vanishes. But in our set-up we need to consider the equation for
$\theta_{DM}$ as well since there is non-zero momentum transfer in the
DM frame. We have checked that in the limit $\wdm=0,\csdm=0,\csde=1$,
\ie in the non-interacting scenario, this framing of equations
provides the same result as in the standard synchronous gauge set-up.

It is now straightforward to verify that the above set of
perturbation equations represent a broad class of cosmological models under
consideration.  They readily boil down to the 6-parameter $\lam$CDM
and non-interacting $w_z$CDM, modified gravity or warm dark matter models with the
following choice of parameters:
\begin{itemize}
\item
$\wdm=0, \wde =-1, \csdm=0, \csde=1$ ~($\lam$CDM)
\item
$\wdm=0, \wde < -1$ (phantom) or $>-1$ (non-phantom), $\csdm=0, \csde=1$
or $\neq 1$ ~(depending on non-interacting $w_z$CDM or modified
gravity models)
\item
$\wdm \neq 0, \wde =-1, \csdm=0, \csde=1$ ~($\lam$WDM)
\item
$\wdm \neq 0, \wde < -1$ or $>-1$, $\csdm=0$ or $\neq 0$ , $\csde=1$
or $\neq 1$ ~(for more complicated warm dark matter models, such as $w_z$WDM)
\end{itemize}

Thus, in a nutshell, we have in our hand a set of
background and perturbation equations for a wide class of cosmological
models in terms of the {\em effective} equations of state and {\em
  effective} sound speeds. Constraining these effective parameters
from data in turn results in studying pros and cons of different class
of cosmological models in this framework. As already
stated, in the rest of the article we are going to primarily address
two major tensions of modern cosmology, namely, the values of $H_0$
and $\sigma_8$ from different low and high redshift data, using the framework described above.

\section{Methodology}\label{sec:data}

We may now test our model-independent framework against currently
available data. Many different cosmological observations are sensitive
to the dark sector. To constrain different class of cosmological
models, both background expansion data and perturbative data may be
utilized. The primary goal in this work is to investigate whether the
inconsistencies in the low and high redshift data can be resolved in
{\em any} class of the cosmological models using this
model-independent framework.  We concentrate on the following
datasets:
\begin{itemize}
\item
CMB: Planck TT and low-$l$ data from the Planck 2015 data release
\citep{planck15}.
\item
Galaxy BAO: Measurements from 6dFGS at $z=0.106$ and MGS at $z=0.15$
from SDSS, as well as the CMASS and LOWz samples from BOSS DR12 at
$z=0.38,0.51$ and $0.61$ \citep{alam16_sdss12}.
\item
SNeIa: SNe Ia data from Joint Light curve Analysis of SDSS-II and
SNLS3 \citep{bet14_jla}.\\
\item
$H_0$: Recent direct measurement of the Hubble constant
\citep{riess16_h0}, which provides a value of $H_0=73.24\pm 1.74$
km/s/Mpc.\\
\end{itemize}	

The combination of datasets outlined above is neither exhaustive nor
complete, and other works are available which provide somewhat
different takes on some of these datasets. For example, direct
measurements of $H_0$ are subject to various tensions. The early HST
Cepheid+SNe based estimate from \citep{riess11_h0} gives $H_0 = 73.8
\pm 2.4$ km/s/Mpc. The same Cepheid data have been analyzed in
\citep{efst14_h0} using revised geometric maser distance to NGC
4258. Using NGC 4258 as a distance anchor, they find $H_0 = 70.6 \pm
3.3$ km/s/Mpc. The more recent paper, \citep{riess16_h0}, obtains a
$2.4\%$ determination of the Hubble Constant at $H_0 = 73.24 \pm 1.74$
km/s/Mpc combining the anchor NGC 4258, Milky Way and LMC
Cepheids. The Milky Way Cepheid solutions for $H_0$ may be unstable
\citep{efst14_h0}, which could go some way in explaining this
inconsistency. However, recent strong lensing observations
\citep{bonv16_h0}, also give the slightly higher value of $H_0 =
71.9^{+2.4}_{-3.0}$ km/s/Mpc. On the other hand, Hubble parameter
measurements from SNe and red giant halo populations,
\citep{tamm13_h0} give $H_0 = 63.7 \pm 2.3$ km/s/Mpc. A recent Hubble
parameter measurement by \citet{ratra16_h0} prefers a value of $H_0 =
68.3^{+2.7}_{-2.6}$ km/s/Mpc, which is more in line with the Planck
results. The most recent SDSS DR12 BAO data \citep{alam16_sdss12} also
appears to favour a somewhat lower value of $H_0= 67.8 \pm 1.2$
km/s/Mpc. Thus as yet there is no clear consensus about the value of
$H_0$. We have chosen to use the result from \citet{riess16_h0} (R16)
since this is the latest direct measurement of $H_0$, and it is
clearly in tension with CMB.

Similarly, although cluster counts for X-ray selected clusters from
REFLEX-II provide a lower value of $\sigma_8 \sqrt{\omt/0.3}= 0.745\pm
0.039$ \citep{bohr14_s8} compared to Planck, an analysis of cluster
counts of X-ray selected clusters by the WtG collaboration,
incorporating the WtG weak lensing mass calibration, finds $\sigma_8
\sqrt{\omt/0.3} = 0.81 \pm 0.03$, \citep{mantz15_s8}, in better
agreement with the Planck CMB results of $\sigma_8 \sqrt{\omt/0.3} =
0.851\pm 0.013$. This discrepancy within cluster observations may be
due to mass calibration biases or biases in the assumed scaling
relations for SZ selected clusters as compared to X-ray selected
clusters. As in the case of $H_0$, here too we shall compare the
$\sigma_8$ obtained from our analysis with that from the more
exhaustive dataset \citep{bohr14_s8}, which is in tension with Planck,
to see if interaction in the dark sector may alleviate this tension.

Within the BAO datasets, the \lya BAO results are in more than
$2\sigma$ tension with the low redshift galaxy BAO results, and are
plagued by various systematics \citep{aub14_bao}, also the SDSS DR12
for these data has not yet been released, hence we leave the \lya data
out of our analysis at present, using the galaxy BAO data only.

To determine the likelihoods for our parameters of interest, we
perform a Monte Carlo Markov Chain analysis with CosmoMC using a
modified version of CAMB. Assuming a flat FRW universe, we may vary
following cosmological parameters: the physical baryon and DM
densities today ($\Omega_bh^2$ and $\Omega_ch^2$), angular size of the
last scattering surface ($\theta$), optical depth due to reionization
($\tau$), amplitude of the primordial power spectrum ($A_s$), scalar
spectral index ($n_s$), effective EoS of DE ($\wde$, which can be
further parameterized by its value today $w_0$, and its rate of change
over the scale factor $w_a$), effective EoS of DM ($\wdm$), effective
sound speed of DE ($\csde$) and effective sound speed of DM ($\csdm$).
	
Therefore, in addition to the standard $\lam$CDM parameters, we now
need to constrain the effective parameters $\lbrace \wde, \wdm, \csde,
\csdm \rbrace$. For the dark energy equation of state, we use the
well-known model-independent Chevallier-Polarski-Linder
parameterization \citep{cp01, lin03}, which takes into account a wide
class of dark energy models (and may represent the effective dark
energy for interacting models in our formalism) and is represented by
\beq
\wde=w_0 + w_a(1-a) \,\,,
\eeq
One may wonder if the above CPL parametrization, that is usually employed for
non-interacting dark energy models, can be used in this generalised scenario. We should clarify
that at this point. A  parametrization is a tool to constrain a number of models
from observations. As is well-known, data is not directly sensitive to models, rather it is sensitive to some
parameters that represent the background model(s) via the parametrization. As such CPL is a  considerably good
parametrization that can take into account most of the non-interacting dark energy models.
Since  in our formalism we have made the effective equations of state look like non-interacting,
it can in principle be applied to represent at least these class of models under consideration, even though that encompasses,
intrinsically, interacting DMDE models, among others. 
Nevertheless, as it will turn out in the subsequent section, present datasets constrain $\wdm$ to
pretty close to zero, and hence any interaction as such has to be very tiny. As a result, effectively, the
$\wde$ behaves pretty close to the EoS of non-interacting models. Hence a CPL parametrization for the
class of models under consideration (non-interacting $w_z$CDM, modified gravity, warm dark matter models, or $\lam$CDM) is very much 
a suitable parametrization. The only assumption made  here is that in case of interacting DMDE models,
the interaction has to be really feeble, which is indeed the case so far as observational data is concerned.

Thus the Hubble parameter, representing the expansion history of
the universe, may be written as:
\beq
H(a)= H_0\left[\omr a^{-4} + \omt a^{-3(1+\wdm)}+(1-\omr-\omt) a^{-3(1+w_0+w_a)} e^{-3 w_a(1-a)}\right]^{1/2} \,\,.
\eeq

In this ansatz, the DE EoS may cross the phantom barrier $(w=-1)$ at
some point of its evolution. Typically, single scalar field models of
dark energy cannot have such a phantom crossing since the velocity
component of the perturbation equations would blow up at $\wde=-1$
(see equation~\ref{eq:pert_v}). It is possible to have such a phantom
crossing in models with multiple scalar fields representing dark
energy \citep{hu08_ppf}. For this work, we limit ourselves to the
simpler single scalar field or modified gravity scenarios and study
phantom and non-phantom behaviour separately. We use the priors $w_0
\in [-1,-0.33]$ and $[-3,-1]$ for non-phantom and phantom regimes
respectively; and $w_a \in [-2,2]$, $\csde \in [0,2]$. We do not
attempt to vary sound speed of DM as it is very tightly constrained by
the available data, and keep it fixed to zero, as expected for
standard cold dark matter. We note here that the parameters varied
here, \ie $w_0, w_a, \wdm, \csde$ are all effective parameters for DM
and DE, which implicitly contain the interaction, if any, between DM
and DE.  Data being sensitive only to the effective parameters, the
true values of $w_{DE}, w_{DM}, c^2_{s, DE}$ are not directly seen by
the observables. The presence and nature of interaction, if any,
between DM and DE can be surmised from the deviation of above
effective values from the standard $\lam$CDM or $w_0$, $w_a$+CDM
parameter values. A non-zero value of $\wdm$, for example, could
either signal a departure from CDM (e.g., warm dark matter models), or
the presence of interaction between CDM and DE. Since available data
strongly constrains the ``coldness'' of dark matter, we interpret any
departure from $\wdm=0$ as the possibility of interaction within the
dark sector.

	
\section{Results, Comparison and Discussions}\label{sec:res}

\subsection{Phantom EoS}

\begin{figure*}
\includegraphics[width=0.9\linewidth]{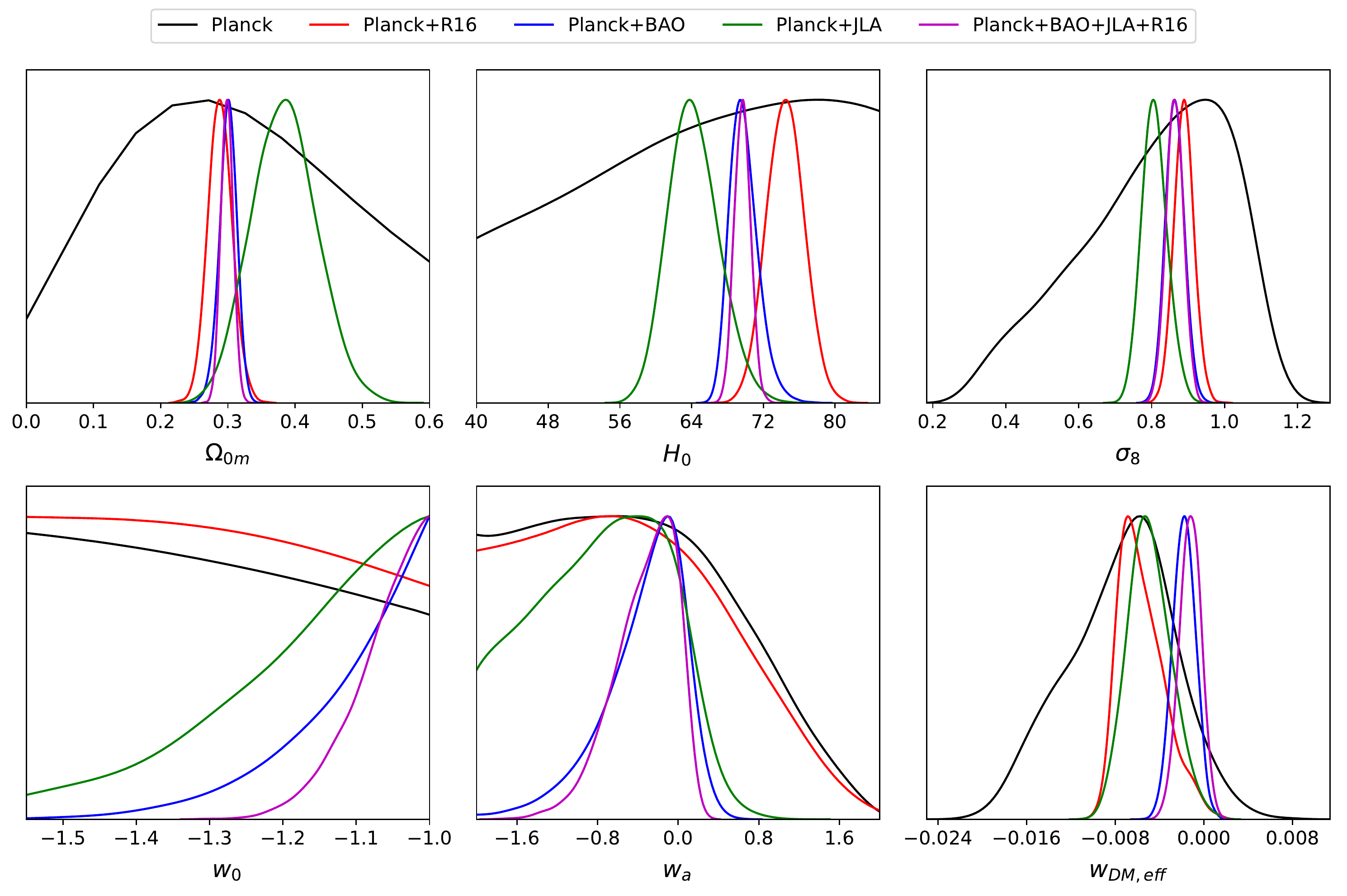} \\
\caption{\footnotesize
Likelihoods in $\omt, H_0, \sigma_8, w_0, w_a, \wdm$ for cosmological
reconstruction using Planck (black lines), Planck+R16 (red lines),
Planck+BSH (blue lines), for phantom (non)interacting models under consideration.}
\label{fig:p_1D}
\end{figure*}

\begin{figure*}
\centering
$\begin{array}{ccc}
\includegraphics[width=0.32\linewidth]{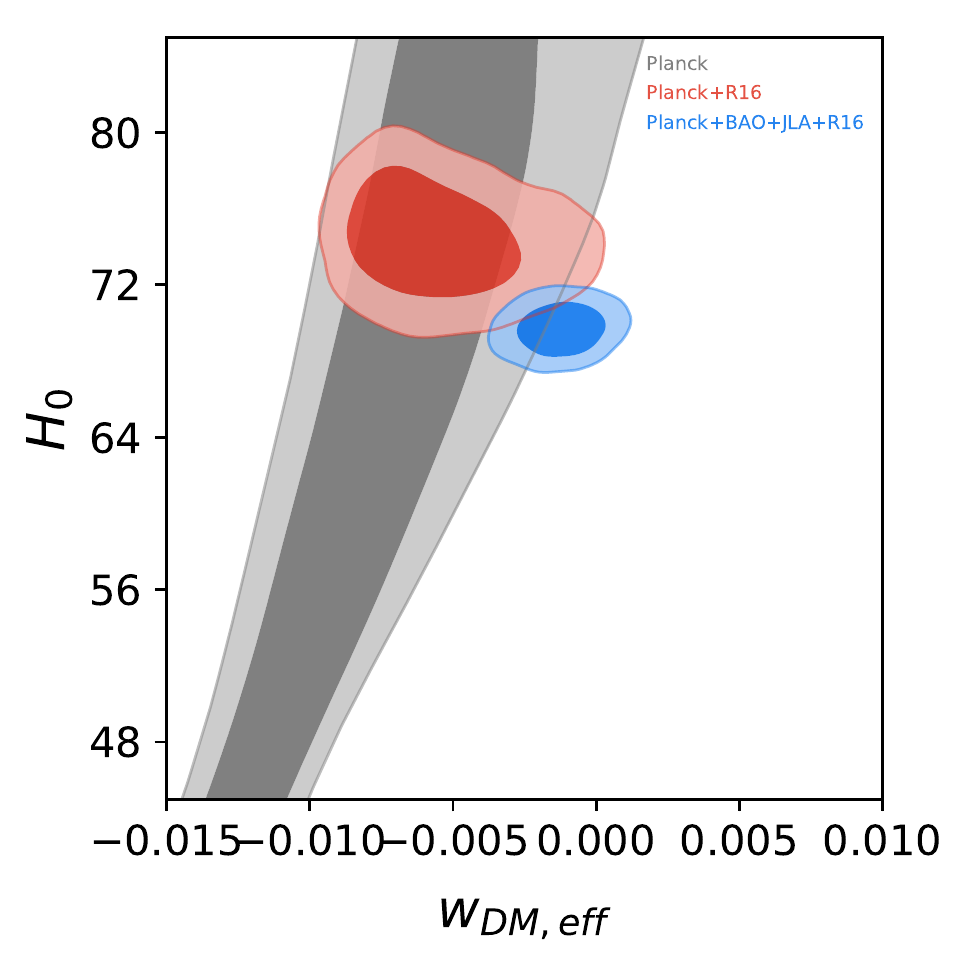} &
\includegraphics[width=0.32\linewidth]{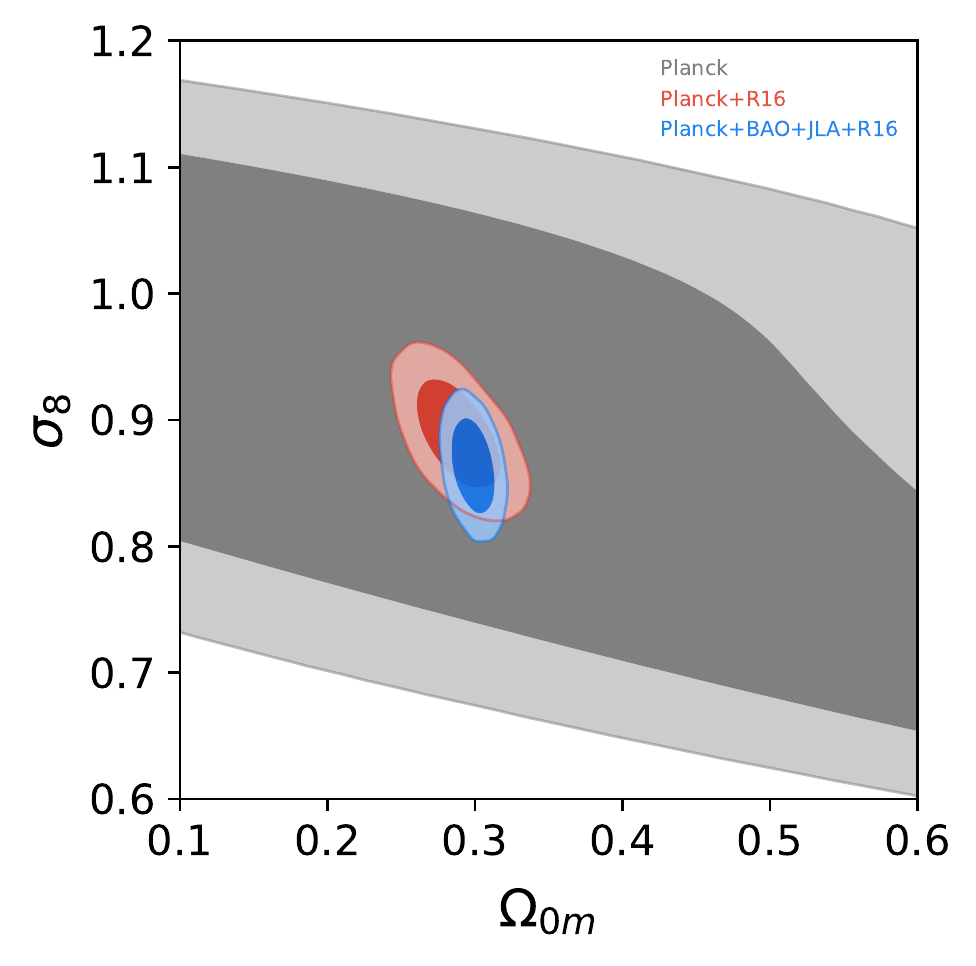} &
\includegraphics[width=0.32\linewidth]{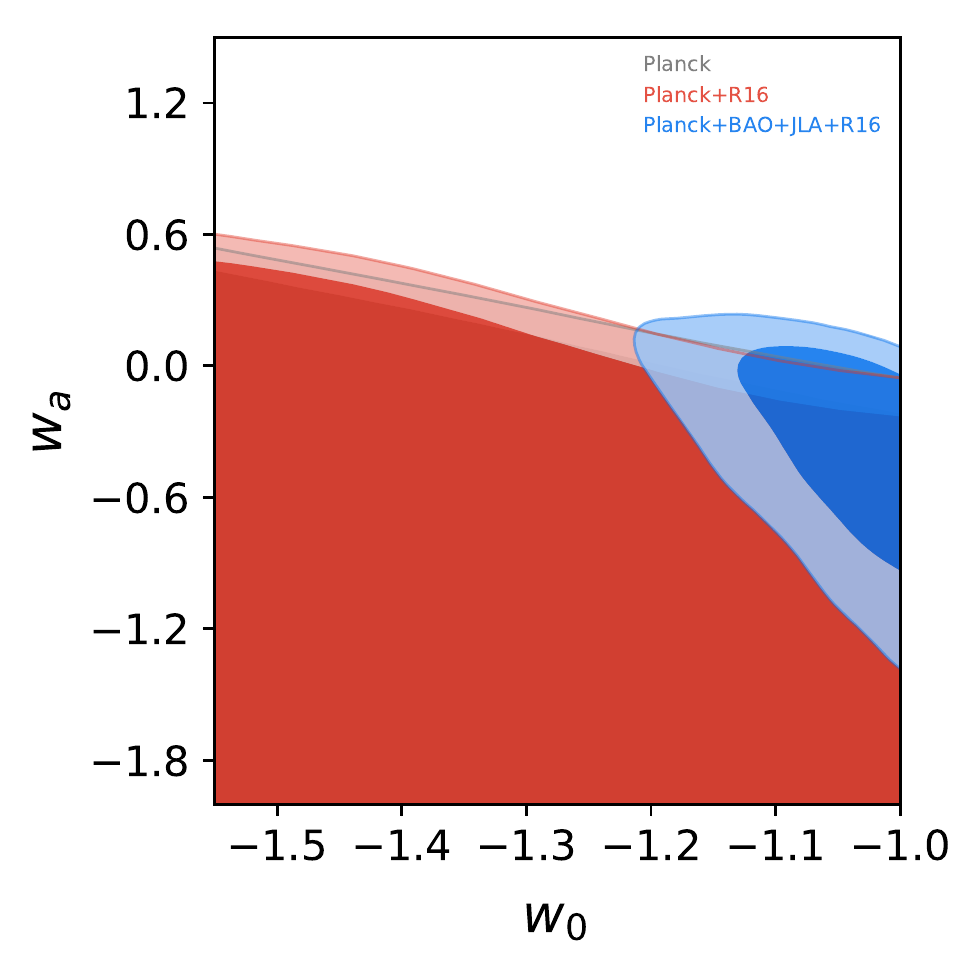} \\
\mbox{(a)} & \mbox{(b)}  & \mbox{(c)}
\end{array}$
\caption{\footnotesize
$1,2\sigma$ confidence levels in the $H_0-\wdm$ (left panel),
$\omt-\sigma_8$ (middle panel), and $w_0-w_a$ (right panel) parameter
spaces using Planck (grey), Planck+R16 (red), Planck+BSH (blue), for
phantom (non)interacting models under consideration.}
\label{fig:p_2D}
\end{figure*}

As pointed out in the last section, we will deal with phantom and
non-phantom cases separately.  We first show the results for the
phantom (non)interacting DMDE models, \ie models with $\wde \leq
-1$. The parameters of interest are $\omt, H_0, \sigma_8, w_0, w_a,
\wdm, \csde$. We wish to see if opening up the parameter space helps
ease the tension in $H_0$ as well as that in $\omt, \sigma_8$. We will
also attempt to understand the effect of different datasets on the
individual parameters, and hence on different classes of models.  
We reiterate here that when we say 'models' here, we have in our mind the usual $\lam$CDM,
non-interacting $w_z$CDM, modified gravity, warm dark matter models
as well as a  class of interacting DMDE models that can be represented 
in this theoretical framework described in Section \ref{sec:setup}.
This
will in turn constrain the dark matter and dark energy equations of
state directly from observations for these wide class of theoretical models.  Firstly, we find that the results are quite insensitive to
$\csde$, freeing up this parameter has little effect on the
constraints of the other parameters, and the parameter itself remains
fairly unconstrained. We therefore keep it fixed for primary analysis
at the scalar field value of $\csde=1$.  Fig~\ref{fig:p_1D} shows the
likelihoods for the remaining parameters using (i) only Planck data,
(ii) Planck with R16 $H_0$ measurement, (iii) Planck with BAO data,
(iv) Planck with SNe data, and (v) Planck with R16 $H_0$ + BAO + SNe
Type Ia data (BSH).

We see that the Planck data by itself (black lines) does not have very
strong constraining power on the individual parameters.  With CMB
alone, $H_0, \omt, \wdm, w_0, w_a$ are all fairly unconstrained.  The
underlying reason is as follows: since we constrain $\delta_{DM}$ or
its function, and $H_0, \wdm$ and $\omt$ enter the perturbation
equations as $\wdm h$ and $\omt h^2$, therefore, although $\wdm h$ and
$\omt h^2$ are constrained quite strongly, $H_0, \wdm$ and $\omt$ are
individually unconstrained since one can always increase one parameter
and decrease another to achieve the same constraint for the
combinations.  The effective dark energy parameters $w_0$ and $w_a$
enter indirectly through $\delta_{DE}$ and therefore metric
perturbations, thus, they or any function of them is not strongly
constrained by CMB.

With the addition of the $H_0$ measurements, CMB+R16 tightens up
constraints on $H_0$ and therefore on $\omt, \wdm$, and consequently
on $\sigma_8$ (red lines), but provides no further constraining power
for the dark energy parameters.  As above, this can be understood as:
$w_0$ and $\wdm$ enter expansion history similarly. However, the
difference between them appear in perturbations. CMB constrains $\omt
h^2$ and $\wdm h$, therefore if $H_0$ is provided a fixed range from
R16, $\omt$ and $\wdm$ also get confined to a fixed narrow range. The
same effect is not seen on $w_0$ or $w_a$ because these or any
functions of them are weakly constrained by CMB.

The addition of BAO to Planck data brings the matter density to $\omt
\sim 0.3$ which is slightly higher and with narrower errors than the
result for Planck+R16, but it chooses an $H_0$ noticeably lower than
that favoured by R16, and also a higher $\wdm$ (blue lines). In
addition it also provides some constraints on $w_0, w_a$. BAO measured
either the Hubble parameter or its integral in the form of the angular
diameter distance, and from these it tends to put strongest
constraints on $\omt$, and weaker constraints on the other parameters
such as $H_0, \wdm, w_0, w_a$. Adding these new constraints to CMB, we
are able to break the degeneracy between $\omt , H_0$ and $\wdm,
H_0$. BAO by itself would allow degeneracy between $\wdm$ and $w_0$,
and between $\wdm$ and $H_0$ as well, this degeneracy is broken by
constraints from CMB on $\wdm h, \ \omt h^2$. Once $\omt, H_0, \wdm$
are constrained, the remaining parameters $w_0, w_a$ get constrained
as well. $w_a$ has the weakest constraint since it enters the equation
for $H(z)$ to the second order.

Adding the JLA SNeIa to Planck narrows down the constraints like BAO,
but in a different direction. In this case, $\omt$ is moved to a
higher value than that for either of the two previous cases, and $H_0$
to a lower value (green lines). $\wdm$ is at about the same region as
that for Planck+$H_0$. The DE parameters are constrained as well, but
less than that in the case for BAO. In this case, we know that JLA+CMB
tends to prefer non-phantom DE \citep{bet14_jla} with $\omt \simeq 0.3$ in
the non-interacting case. Here we are adding a new parameter $\wdm$,
and constraining the DE parameters to the phantom regime, forcing
$\wde \leq -1$.  The data may compensate for phantom DE by choosing
either (i) $\omt > 0.3$, $\wdm \lleq 0$, or (ii) $\omt < 0.3$, $\wdm
\ggeq 0$. Since the CMB data prefers to keep the new parameter $\wdm <
0$, Planck+JLA therefore pushes $\omt$ to a higher value and
consequently $H_0$ to a lower value. The DE parameters are less
constrained than BAO because BAO measures $H(z)$ while SNe data
measures the magnitude which is related to $H(z)$ by an integral and
logarithm, thereby reducing its constraining power.

Adding all the datasets together, naturally the constraints are at
their narrowest (purple lines), however, given the inconsistencies
between the different datasets, the results are not necessarily
commensurate with those for the separate datasets. For example,
Planck+R16 obtained a high $H_0$, but due to the effect of SNe and
BAO, Planck+BSH reduces $H_0$. Thus though the tension between CMB and
direct $H_0$ is resolved for a slightly negative $\wdm$, Planck+BSH
does not completely agree with the direct $H_0$ measurements.

Fig~\ref{fig:p_2D} shows the $1,2\sigma$ confidence levels in
the $H_0-\wdm$, $\omt-\sigma_8$ and $w_0-w_a$ parameter spaces for the
three different datasets. We see here that the Planck confidence
levels in $H_0$ are very large (grey contours of left panel), mainly
due to the flexibility afforded by the new parameter $\wdm$. For small
negative values of $\wdm$, therefore, $H_0$ from Planck data is
allowed to go up to much larger values than those allowed by
$\lam$CDM, thereby reducing its tension with the direct measurement of
$H_0$ (as evinced from the red contours in the left panel). The
addition of BAO and SNe data however, slightly disfavours non-zero
$\wdm$, and the tension in $H_0$ resumes somewhat. 

Further, due to the freeing up of $H_0$, the $\omt-\sigma_8$ parameter
space is also opened up, with lower values of $\sigma_8$ chosen for
higher values of $\omt$. The Planck results therefore have the
potential to be commensurate with the cluster results, since
$\omt=0.3,\sigma_8=0.75$ falls well within the $1\sigma$ levels (grey
contours of middle panel). However, both BAO and $H_0$ measurements
appear to push the $\sigma_8$ to higher values, mainly because
$\sigma_8$ has a positive correlation with $H_0$, \ie higher the
$H_0$, higher the value of $\sigma_8$. Thus by increasing the value of
$H_0$ to fit BSH, we reduce consistency with cluster results for
$\sigma_8$, since lower $H_0$ and therefore lower $\sigma_8$ is
disfavoured when these datasets are added to Planck (red and blue
contours of left and middle panels).

The effective equation of state of dark energy is constrained
only with the addition of BAO and SNe data: while $w_0 \simeq -1.2$ at
$2\sigma$, the rate of change $w_a$ is allowed a fairly large range,
going down to $w_a \ggeq -1.6$.

\begin{table*}
\caption{\footnotesize
Best-fit and $1\sigma$ values for $\omt, H_0, \sigma_8, w_0, w_a,
\wdm, \csde$ and best-fit $\chi^2$ for phantom {\bf (non)interacting models under consideration}
using Planck, Planck+R16, Planck+BSH. Corresponding values for
$\lam$CDM and CPLCDM are given for comparison. }
\label{tab:p}
\begin{tabular}{c|c|c|c|c|c|c|c|c|c|c}
\hline
Data&Model&$\omt$&$H_0$&$\sigma_8$&$w_0$&$w_a$&$\wdm$&$\csde$&$\chi^2_{\rm bf}$&$\chi^2_{\lam CDM}-\chi^2_{\rm bf}$\\
\hline
&$\lam$CDM&$0.30^{+0.02}_{-0.02}$&$68.1^{+1.2}_{-1.2}$&$0.85^{+0.03}_{-0.02}$&$-1$&$0$&$0$&$1$&$781.07$&$0$\\
Planck&CPLCDM&$0.19^{+0.02}_{-0.04}$&$88.4^{+11.6}_{-3.7}$&$1.02^{+0.08}_{-0.06}$&$-1.5^{+0.3}_{-0.3}$&$-0.13^{+0.27}_{-0.03}$&$0$&$1$&$779.83$&$-1.24$\\
&$+\wdm$&$0.62^{+0.32}_{-0.59}$&$66.7^{+32.0}_{-11.1}$&$0.80^{+0.27}_{-0.13}$&$-2.0^{+1.0}_{-1.0}$&$-0.45^{+0.48}_{-1.50}$&$-0.0075^{+0.005}_{-0.004}$&$1$&$778.26$&$-2.81$\\
&$+\csde$&$0.68^{+0.32}_{-0.66}$&$64.9^{+31.7}_{-13.5}$&$0.79^{+0.28}_{-0.14}$&$-2.0^{+1.0}_{-1.0}$&$-0.42^{+0.49}_{-1.58}$&$-0.0078^{+0.005}_{-0.004}$&$1.03^{+0.84}_{-0.45}$&$778.88$&$-2.19$\\
\hline
&$\lam$CDM&$0.29^{+0.01}_{-0.01}$&$69.7^{+1.0}_{-1.0}$&$0.86^{+0.02}_{-0.02}$&$-1$&$0$&$0$&$1$&$786.66$&$0$\\
Planck&CPLCDM&$0.26^{+0.01}_{-0.01}$&$74.0^{+1.7}_{-1.7}$&$0.90^{+0.02}_{-0.03}$&$-1.1^{+0.1}_{-0.1}$&$-0.27^{+0.46}_{-0.26}$&$0$&$1$&$782.02$&$-4.64$\\
+R16&$+\wdm$&$0.29^{+0.02}_{-0.02}$&$74.5^{+2.1}_{-2.1}$&$0.88^{+0.03}_{-0.03}$&$-2.0^{+1.0}_{-1.0}$&$-0.96^{+1.10}_{-1.50}$&$-0.005^{+0.001}_{-0.003}$&$1$&$777.65$&$-9.01$\\
&$+\csde$&$0.29^{+0.02}_{-0.02}$&$74.5^{+2.1}_{-2.2}$&$0.89^{+0.02}_{-0.02}$&$-2.0^{+1.0}_{-1.0}$&$-0.94^{+1.05}_{-1.65}$&$-0.005^{+0.001}_{-0.002}$&$1.03^{+0.96}_{-0.34}$&$780.19$&$-6.47$\\
\hline
&$\lam$CDM&$0.30^{+0.01}_{-0.01}$&$68.5^{+0.6}_{-0.6}$&$0.86^{+0.02}_{-0.02}$&$-1$&$0$&$0$&$1$&$1490.66$&$0$\\
Planck&CPLCDM&$0.29^{+0.01}_{-0.01}$&$69.8^{+1.0}_{-1.0}$&$0.87^{+0.02}_{-0.02}$&$-1.05^{+0.05}_{-0.01}$&$-0.15^{+0.21}_{-1.10}$&$0$&$1$&$1490.29$&$-0.37$\\
+BSH&$+\wdm$&$0.30^{+0.01}_{-0.01}$&$69.7^{+1.0}_{-1.0}$&$0.86^{+0.02}_{-0.02}$&$-1.06^{+0.06}_{-0.01}$&$-0.33^{+0.39}_{-0.19}$&$-0.0012^{+0.001}_{-0.001}$&$1$&$1488.14$&$-2.52$\\
&$+\csde$&$0.30^{+0.01}_{-0.01}$&$69.7^{+1.0}_{-1.0}$&$0.86^{+0.02}_{-0.02}$&$-1.06^{+0.06}_{-0.01}$&$-0.34^{+0.40}_{-0.18}$&$-0.0012^{+0.001}_{-0.001}$&$1.02^{+0.98}_{-1.02}$&$1488.83$&$-1.83$\\
\hline
\end{tabular}
\end{table*}

We are now in a position to make use of these results to compare  among different types of  models under consideration, some of whom
have  less number of free
parameters (namely, $\lam$CDM or non-interacting $w_z$CDM with
phantom-like behaviour using CPL ansatz for $w_{DE}$ again). We can readily do so by comparing
the best-fit, $1\sigma$ values for the different parameters, as well
as the best-fit $\chi^2$ in the table~\ref{tab:p}. We see that for
Planck data only, the $\chi^2$ for CPLCDM is slightly better than that
for $\lam$CDM, although not at a significance where it could be
comprehensively claimed that phantom variable dark energy models are
favoured over $\lam$CDM. Introducing $\wdm$, which is equivalent to
introducing a coupling between DM and DE (or introducing work dark matter models), does improve the $\chi^2$
over $\lam$CDM slightly more in the phantom case. The addition of the
$\csde$ parameter, on the contrary, degrades the $\chi^2$ very
slightly, possibly because the parameter space now has too many
degeneracies, thus reducing the constraining abilities of the data. We
also see that for just Planck data, CPLCDM may allow much higher
values of $H_0$ than that for $\lam$CDM, but for lower values of
$\omt$. In fact, the value chosen for $H_0$ is so high that it is now
incommensurate with R16, but from the higher end, with a lower $\omt$
to boot. Thus we cannot achieve consistency between Planck and R16 by
simply allowing dynamical DE in the phantom regime. When BAO and SNe
are added, $\omt$ increases, reducing the value of $H_0$ again to
$\lam$CDM levels, so putting all the data together results in
constraints very similar to that for $\lam$CDM, albeit with a slightly
better $\chi^2$. The addition of $\wdm$ opens up the $H_0$ parameter
space, and a much larger range of values are allowed for both $H_0$
and $\omt$, for even a slightly non-zero value of $\wdm$. Thus
consistency with R16 is achieved with $\omt \simeq 0.3$. Once again,
however, the addition of BAO and SNe constrains $\wdm \simeq 0$,
bringing the value of $H_0$ down slightly, although it is still higher
than that for $\lam$CDM. For $\sigma_8$, we find that the Planck data
by itself does allow for a lower $\sigma_8$ for reasonable values of
$\omt$. However, the addition of R16, or of BSH, increases $\sigma_8$
in response to the corresponding increase in $H_0$.  The $\sigma_8$
parameter may take on lower values for just Planck data, but it still
appears to favour higher values when all data is taken together, so
the tension with cluster data remains unresolved.

So, in a nutshell, the results for phantom case can be summarized as
below:
\begin{itemize} 
\item
$H_0$ tension: 
\begin{itemize}
\item
R16: gives high $H_0$.
\item  
CMB: $\lam$CDM prefers low $H_0$, non-interacting CPLCDM has too high
$H_0$ and too low $\omt$. In comparison, in these class of interacting CPLCDM  or warm dark matter models, $\omt$
is fairly unconstrained, hence although a positive correlation between
$H_0$ and $\omt$ remains, it is possible to obtain high $H_0$ to R16
levels for a large range of $\omt$, which is a distinct improvement
over both $\lam$CDM and CPLCDM.
\item
BAO: $\lam$CDM and non-interacting CPLCDM both prefer low $H_0$ (or
possibly high $\omt$).  Interacting CPLCDM too appears to prefer
slightly low $H_0$, but it is more in line with the R16 value,
therefore the tension between $H_0$ and CMB can be partially resolved
even after the addition of BAO data.
\item
JLA: Since this dataset prefers non-phantom dark energy, and Planck
prefers negative $\wdm$, addition of this dataset can only serve to
increase $\omt$ and therefore decrease $H_0$, thus exacerbating the
tension with the high value of $H_0$ obtained by R16.
\item
Therefore for the CPLCDM case, tension between R16 and Planck is
resolved for reasonable values of $\omt$, which is not possible for
both $\lam$CDM and CPLCDM.  However, the tension between BAO and $H_0$
is only partially resolved, and addition of SNe data makes the tension
with $H_0$ reappear.
\end{itemize}
\item $\sigma_8$ tension:
\begin{itemize}
\item
Clusters prefer low $\sigma_8$.
\item  
CMB: $\lam$CDM prefers low $H_0$, but not low enough to allow cluster
$\sigma_8$.  For non-interacting CPLCDM, addition of CPL causes
opening up of parameter space with higher $H_0$ and $\sigma_8$. So one
cannot get low $\sigma_8$, the tension becomes worse.  However,  for these class of interacting CPLCDM  or warm dark matter models, 
addition of $\wdm$ causes opening up of parameter
space, for both higher and lower $H_0$ and $\sigma_8$, therefore
tension with clusters is resolved if we allow lower $H_0$.
\item
CMB+R16: For $\lam$CDM and CPLCDM, there is no improvement over the
CMB result.  For interacting CPLCDM as well, due to positive
correlation between $\sigma_8$ and $H_0$, as R16 prefers higher values
of $H_0$, higher $\sigma_8$ is preferred.
\item
CMB+BSH: For $\lam$CDM and CPLCDM we see no improvement over the CMB
result, higher $H_0$ means higher $\sigma_8$.  However, for
interacting CPLCDM, slightly lower $H_0$ is preferred (due to the
presence of BAO and SNe data), therefore slightly lower $\sigma_8$ is
also allowed, although not enough to resolve tension with
clusters. This, however, comes at the cost of inconsistency with the
R16 $H_0$ measurements.
\end{itemize}
\end{itemize}

\begin{figure*}
\centering
$\begin{array}{cc}
\includegraphics[width=0.32\linewidth]{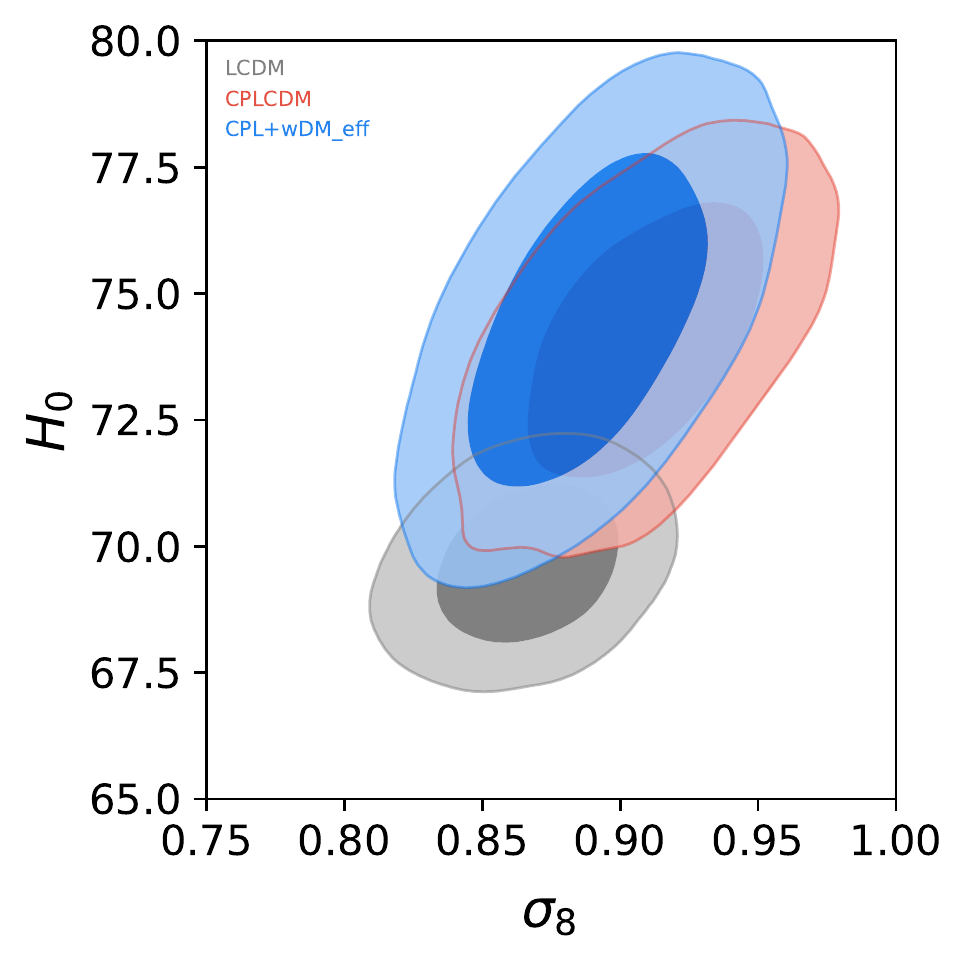} &
\includegraphics[width=0.32\linewidth]{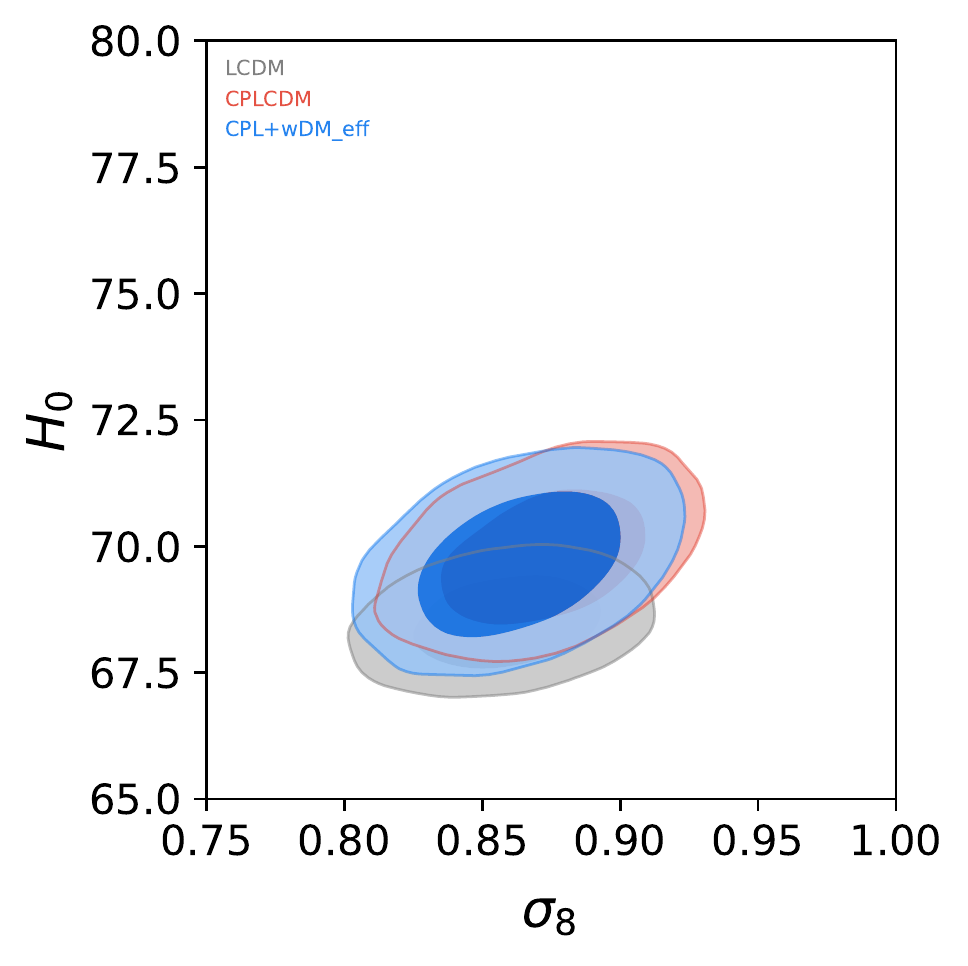} \\
\mbox{(a)} & \mbox{(b)}
\end{array}$
\caption{\footnotesize
$1,2\sigma$ confidence levels in the $H_0-\sigma_8$ parameter spaces
using Planck+R16 (left panel), and Planck+BSH (right panel), for (i)
$\Lambda$CDM (grey), (ii) CPLCDM (red) and (iii) (non)phantom interacting
 (blue) models under consideration.}
\label{fig:p_sig_H0}
\end{figure*}

It transpires from the above discussion that there appears to be a
positive correlation between $H_0$ and $\sigma_8$, no matter whether
we choose $\lam$CDM, non-interacting $w_z$CDM, warm dark matter or a wide class of interacting DMDE as
the cosmological model.  In order to depict this positive correlation
in a more concrete language, we have plotted the $1,2\sigma$
confidence levels in the $H_0 - \sigma_8$ parameter space in
Fig. \ref{fig:p_sig_H0}. To compare among different datasets, we show
the confidence levels for Planck+R16 in the left panel and Planck+BSH
in the right panel, for (i) $\lam$CDM (grey contours), (ii) phantom
CPLCDM (red contours) and (iii) phantom interacting DMDE (blue
contours) models.  We see that as we free up more parameters, the
correlation becomes more significant in case of Planck+R16 although we
confine the parameter space to comparatively higher values of $H_0$
(due to R16). For Planck+BSH, the correlation is relatively less
apparent due to use of BAO and SNe data which confines the results to
the low $H_0$ space.  Thus, the positive correlation appears to be
generic to CMB data, which persists even after adding the low redshift
datasets, and hence, a higher $H_0$ is simply not consistent with a
low $\sigma_8$, and both the tensions cannot be simultaneously
resolved, at least for a fairly general class of cosmological models using present datasets.

Thus from this section we find that firstly, phantom DE models are
very slightly favoured (or at least not disfavoured) over $\lam$CDM,
and allowing even a very small interaction between DM and DE does
provide an even better fit to the CMB data. Varying the sound speed of
dark energy does not improve the fit. Secondly, we note that the
tension between $H_0$ direct measurement and Planck measurement of
$H_0$ can be eased by introduction of a small, negative $\wdm$. This
implies that a class of interacting dark energy models with energy transfer from
dark energy to dark matter, with a more phantom dark energy EoS, and a
slower rate of redshift of dark matter can resolve this tension. When
all the data is put together, a slightly negative $\wdm$ and slightly
phantom $w_0$ (and negative $w_a$ implying that dark energy was even
more phantom-like in the past) is still favoured over
$\lam$CDM. Therefore, a major success of  our analysis making use of
effective phantom EoS is that it   gives rise to a
consistent $H_0$ for CMB+R16 with a considerably good value of
$\omt$ at least for a class of interacting DMDE models. Thus, these class of  models with effective phantom EoS get a
slight edge over the others so far as present data are concerned.
Lastly, the bottom-line for the $\sigma_8$ tension is that
non-interacting $w_z$CDM cannot resolve tension between clusters and
Planck CDM. These type of interacting CPLCDM can resolve tension if lower $H_0$
allowed. If, however, $H_0$ is high, we cannot get low $\sigma_8$ from
CMB, therefore tension of CMB with $H_0$ and $\sigma_8$ can be
resolved separately, but not together.
However, since the effective EoS for dark matter $\wdm$ prefers slightly negative value,
warm dark matter models are not that favoured compared to these class of interacting models.

We remind the curious reader that the EoS for dark matter is the
effective EoS even though the actual EoS may indicate CDM.  An
effective negative EoS for dark matter, as obtained in
Table~\ref{tab:p} may be looked upon as follows.  In a class of interacting
DMDE sector where energy transfer happens from dark energy to dark
matter is slightly preferred.  In this regard, it is interesting to
point out that there exists a well studied model where a simple Yukawa
type interaction between dark matter Fermion and dark energy scalar
$\exp[{\frac{\beta \varphi} {M_{\scriptscriptstyle\rm P}}}] 
\tilde{\psi}_{\scriptscriptstyle\rm DM}\psi_{\scriptscriptstyle\rm DM}$ 
with a runway scalar potential automatically transfer energy to dark matter 
from dark energy with $\beta >0$ \citep{dam94_dmde, amen00_dmde, das06_dmde}.
This type of model with positive $\beta $ can have their origin naturally 
in string theory. Due to this energy intake over Hubble time, dark matter
redshifts slower than $1/a^3$ and as a result acquires an effective
negative equation of state.

\subsection{Non-phantom EoS}

\begin{figure*}
\includegraphics[width=0.9\linewidth]{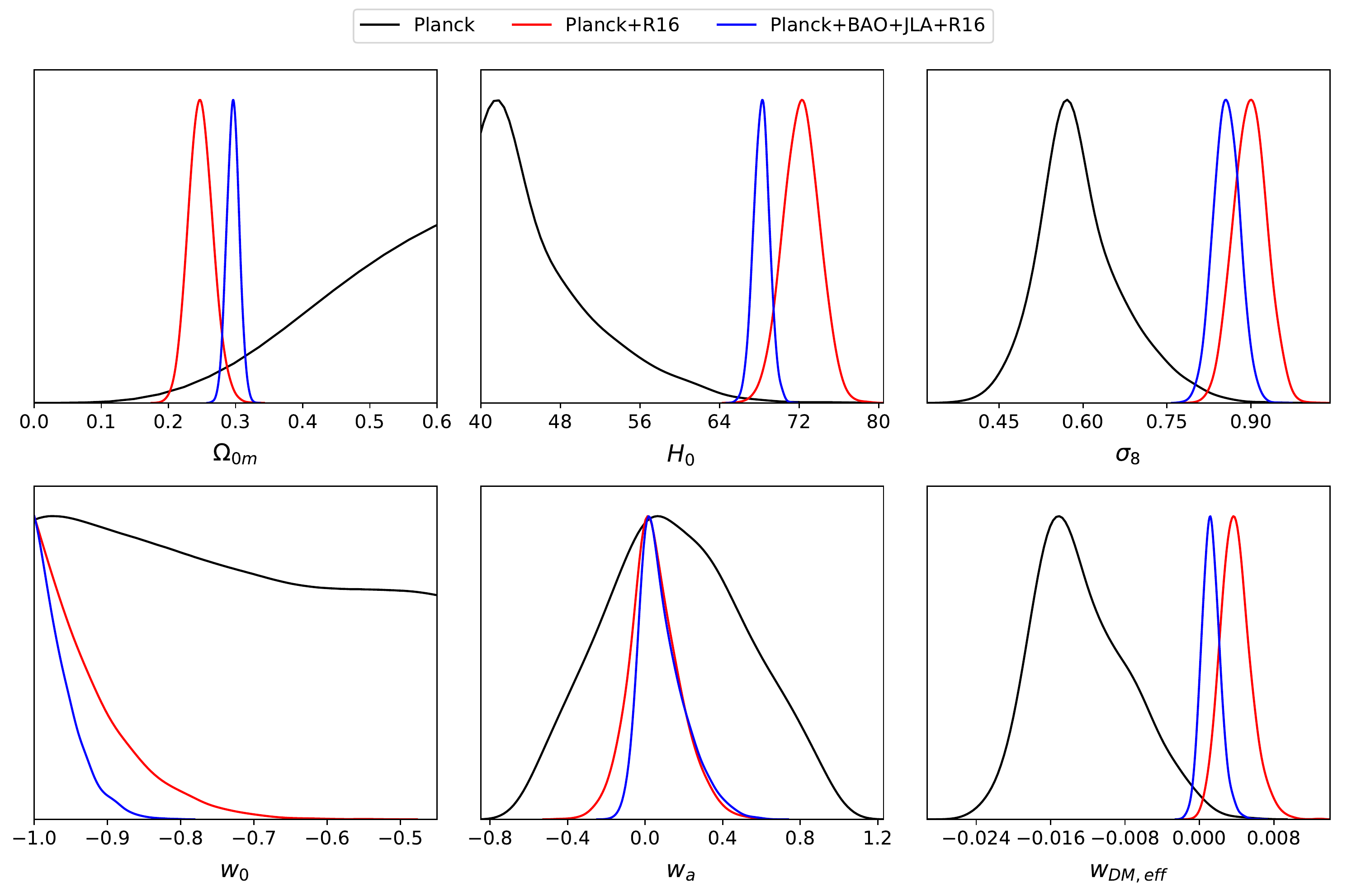} \\
\caption{\footnotesize
Likelihoods in $\omt, H_0, \sigma_8, w_0, w_a, \wdm$ for cosmological
reconstruction using Planck (black lines), Planck+R16 (red lines),
Planck+BSH (blue lines), for non-phantom (non)interacting  models under consideration.}
\label{fig:np_1D}
\end{figure*}

\begin{figure*}
\centering
$\begin{array}{ccc}
\includegraphics[width=0.32\linewidth]{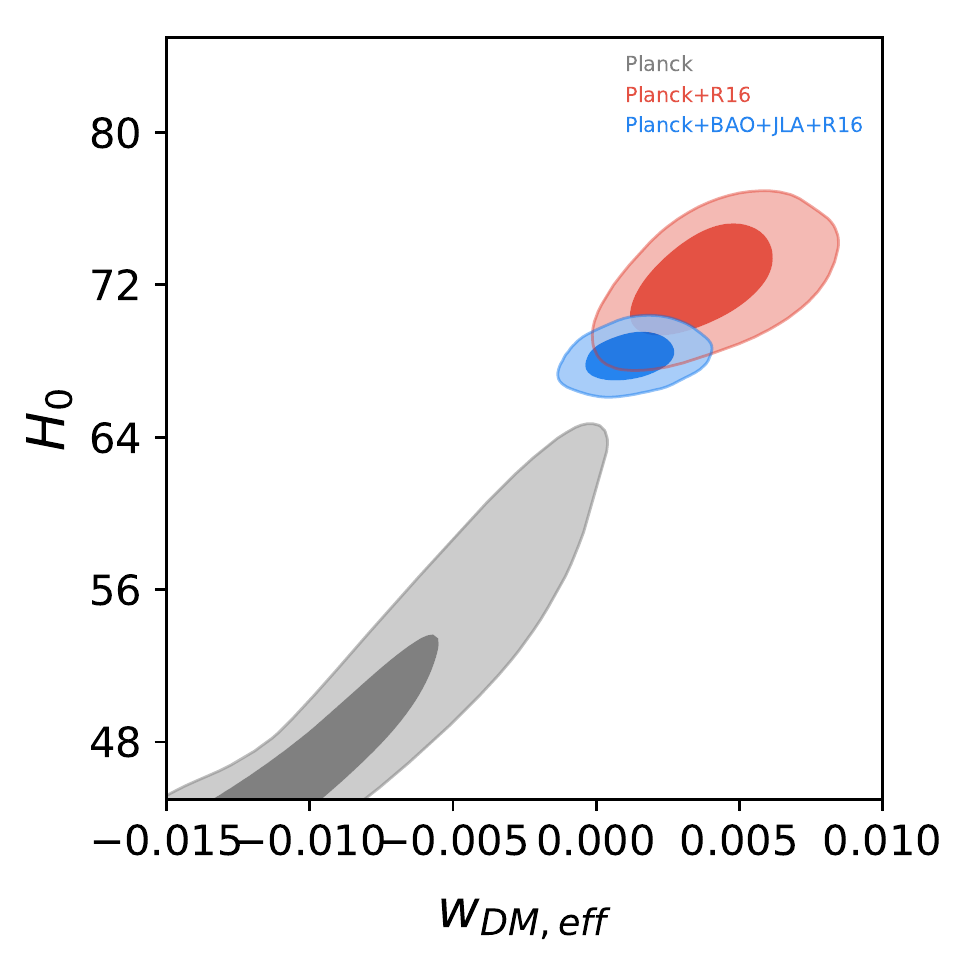} &
\includegraphics[width=0.32\linewidth]{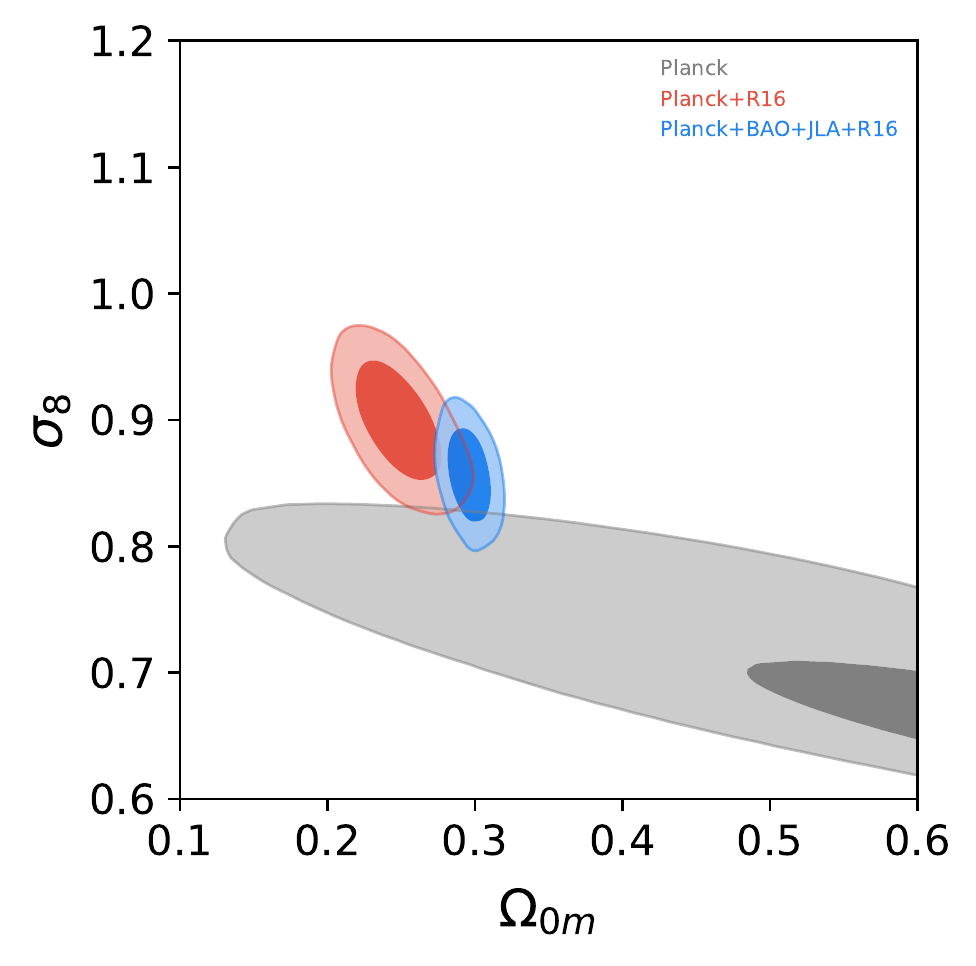} &
\includegraphics[width=0.32\linewidth]{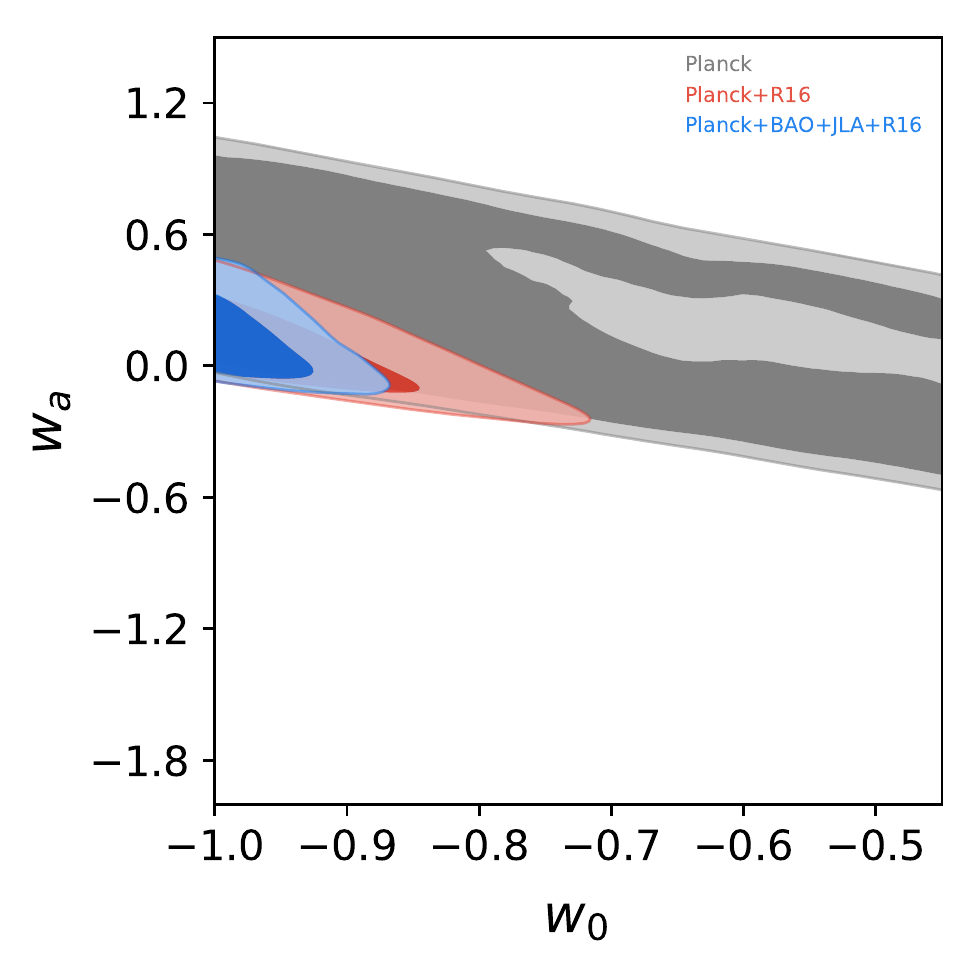} \\
\mbox{(a)} & \mbox{(b)}  & \mbox{(c)}
\end{array}$
\caption{\footnotesize
$1,2\sigma$ confidence levels in the $H_0-\wdm$ (left panel),
$\omt-\sigma_8$ (middle panel), and $w_0-w_a$ (right panel) parameter
spaces using Planck (grey), Planck+R16 (red), Planck+BSH (blue), for
non-phantom (non)interacting DMDE models under consideration.}
\label{fig:np_2D}
\end{figure*}

We now look at the same datasets in the non-phantom \ie $\wde \geq -1$
space for the same class of models, namely, $\lam$CDM, non-interacting CPLCDM, warm dark matter as well as a class of interacting DMDE models. In this case too, $\csde$ has
minimal effect on the results. Figure~\ref{fig:np_1D} shows the
likelihoods for the remaining six parameters. Unlike in the previous
case, Planck data alone (black) shows a preference for much lower
$H_0$ and much higher $\omt$. The parameter $\wdm$ is still negative,
but the likelihoods for $\omt, H_0, \sigma_8, \wdm$ in the case of
Planck all appear to be inconsistent with those for Planck+R16 (red)
and Planck+BSH (blue). This shows that for non-phantom scenario,
Planck CMB results are at odds with those from other
data. Figure~\ref{fig:np_2D} shows the $1,2\sigma$ confidence levels
for $H_0-\wdm$, $\omt-\sigma_8$ and $w_0-w_a$. Whereas in the phantom
case, the extra parameter was liberating both high and low values of
$H_0$, here we see that the $H_0-\wdm$ confidence levels are
inconsistent with those from other data at nearly $2\sigma$. Low
values of matter density are strongly disfavoured, as well as high
values of $\sigma_8$, once again making Planck by itself inconsistent
with other datasets. The equation of state of dark energy appears to
be more constrained than in the phantom case when all data is
considered, leaving very little flexibility. Thus here the tension in
$H_0$ is not resolved as lower values of $H_0$ are so strongly
favoured by Planck, neither is the $\sigma_8$ tension eased.
     
\begin{table*}
\caption{\footnotesize
Best-fit and $1\sigma$ values for $\omt, H_0, \sigma_8, w_0, w_a,
\wdm, \csde$ and best-fit $\chi^2$ for non-phantom (non)interacting  models under consideration
using Planck, Planck+R16, Planck+BSH. Corresponding values for
$\lam$CDM and CPLCDM are given for comparison. }
\label{tab:np}
\begin{tabular}{c|c|c|c|c|c|c|c|c|c|c}
\hline
Data&Model&$\omt$&$H_0$&$\sigma_8$&$w_{0,{\rm eff}}$&$w_{a,{\rm eff}}$&$\wdm$&$\csde$&$\chi^2_{\rm bf}$&$\chi^2_{\lam CDM}-\chi^2_{\rm bf}$\\
\hline
&$\lam$CDM&$0.30^{+0.02}_{-0.02}$&$68.1^{+1.2}_{-1.2}$&$0.85^{+0.03}_{-0.02}$&$-1$&$0$&$0$&$1$&$781.07$&$0$\\
Planck&CPLCDM&$0.37^{+0.03}_{-0.05}$&$62.5^{+4.0}_{-2.7}$&$0.80^{+0.04}_{-0.03}$&$-0.82^{+0.14}_{-0.18}$&$0.03^{+0.22}_{-0.22}$&$0$&$1$&$782.75$&$1.68$\\
&$+\wdm$&$1.06^{+0.31}_{-0.47}$&$44.0^{+4.3}_{-7.5}$&$0.60^{+0.05}_{-0.08}$&$-0.68^{+0.35}_{-0.32}$&$0.16^{+0.36}_{-0.40}$&$-0.012^{+0.004}_{-0.006}$&$1$&$782.63$&$1.56$\\
&$+\csde$&$1.03^{+0.33}_{-0.43}$&$44.5^{+3.7}_{-8.0}$&$0.60^{+0.04}_{-0.09}$&$-0.68^{+0.05}_{-0.35}$&$0.16^{+0.36}_{-0.40}$&$-0.012^{+0.003}_{-0.006}$&$0.98^{+1.02}_{-0.98}$&$780.58$&$-0.49$\\
\hline
&$\lam$CDM&$0.29^{+0.01}_{-0.01}$&$69.7^{+1.0}_{-1.0}$&$0.86^{+0.02}_{-0.02}$&$-1$&$0$&$0$&$1$&$786.66$&$0$\\
Planck&CPLCDM&$0.29^{+0.01}_{-0.01}$&$68.6^{+1.3}_{-1.1}$&$0.85^{+0.02}_{-0.02}$&$-0.97^{+0.01}_{-0.03}$&$0.03^{+0.04}_{-0.06}$&$0$&$1$&$788.97$&$2.31$\\
+R16&$+\wdm$&$0.25^{+0.02}_{-0.02}$&$72.2^{+1.8}_{-1.8}$&$0.89^{+0.03}_{-0.03}$&$-0.92^{+0.01}_{-0.08}$&$0.05^{+0.48}_{-0.13}$&$0.004^{+0.001}_{-0.001}$&$1$&$785.81$&$-0.85$\\
&$+\csde$&$0.25^{+0.02}_{-0.02}$&$72.2^{+1.8}_{-1.8}$&$0.89^{+0.03}_{-0.03}$&$-0.92^{+0.01}_{-0.08}$&$0.05^{+0.10}_{-0.59}$&$0.004^{+0.001}_{-0.002}$&$1.94^{+0.06}_{-1.94}$&$785.73$&$-0.93$\\
\hline
&$\lam$CDM&$0.30^{+0.01}_{-0.01}$&$68.5^{+0.6}_{-0.6}$&$0.86^{+0.02}_{-0.02}$&$-1$&$0$&$0$&$1$&$1490.66$&$0$\\
Planck&CPLCDM&$0.30^{+0.01}_{-0.01}$&$67.8^{+0.7}_{-0.7}$&$0.85^{+0.02}_{-0.02}$&$-0.97^{+0.01}_{-0.03}$&$0.04^{+0.04}_{-0.08}$&$0$&$1$&$1493.36$&$2.70$\\
+BSH&$+\wdm$&$0.30^{+0.01}_{-0.01}$&$68.2^{+0.8}_{-0.8}$&$0.85^{+0.02}_{-0.02}$&$-0.96^{+0.01}_{-0.04}$&$0.10^{+0.07}_{-0.14}$&$0.0012^{+0.001}_{-0.001}$&$1$&$1491.01$&$0.35$\\
&$+\csde$&$0.30^{+0.01}_{-0.01}$&$68.2^{+0.8}_{-0.8}$&$0.86^{+0.02}_{-0.02}$&$-0.96^{+0.01}_{-0.04}$&$0.09^{+0.06}_{-0.14}$&$0.0012^{+0.001}_{-0.001}$&$0.98^{+1.02}_{-0.98}$&$1491.59$&$0.93$\\
\hline
\end{tabular}
\end{table*}

We compare  these results against different models under consideration,  namely, $\lam$CDM, non-interacting CPLCDM,
a class of interacting CPLCDM, and warm dark matter in
table~\ref{tab:np}. In the non-phantom scenario, for all datasets, it
appears that $\lam$CDM has the better $\chi^2$ as compared to CPLCDM,
as well as interacting models. The addition of $\wdm$ improves the
$\chi^2$ slightly from the CPLCDM scenario, but it is still greater
than that of $\lam$CDM. Therefore we may conclude that the
cosmological constant is favoured over non-phantom dark energy models,
even when we include an interaction in the dark sector. As expected,
even with the added parameters, the best-fit values for the standard
parameters $\omt, H_0, \sigma_8$ are pretty close to the $\lam$CDM
values, even the dark energy parameters are close to
$w_0,w_a=-1,0$. When all data is considered, $\wdm$ has a slightly
positive value, but as noted before, this is not statistically
favoured over $\lam$CDM. We note here that the JLA SNe data is
probably the only dataset that favours non-phantom $w_{DE}$ over
phantom $w_{DE}$, but as the other datasets strongly disfavour
non-phantom, the effect of JLA is not felt in these results.
Here also $\wdm$ is still slightly negative, disfavoring warm dark matter models, at least from present datasets.

\begin{figure*}
\centering
$\begin{array}{cc}
\includegraphics[width=0.32\linewidth]{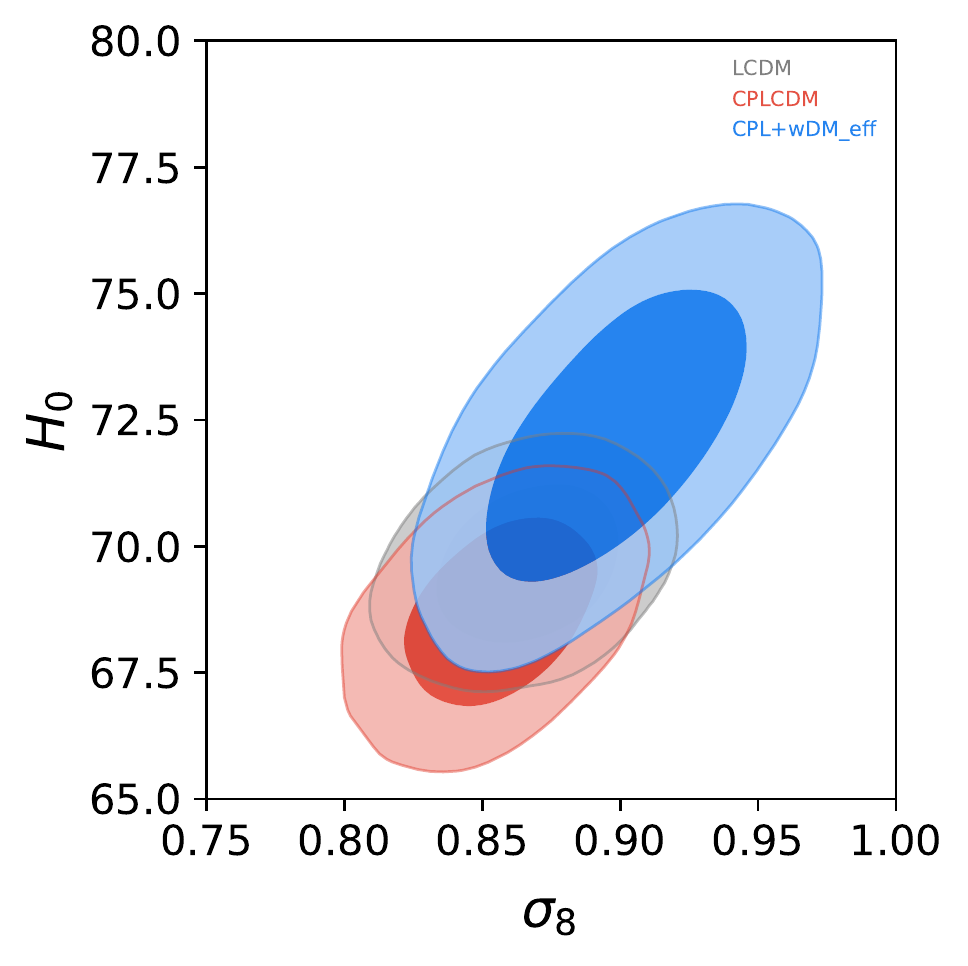} &
\includegraphics[width=0.32\linewidth]{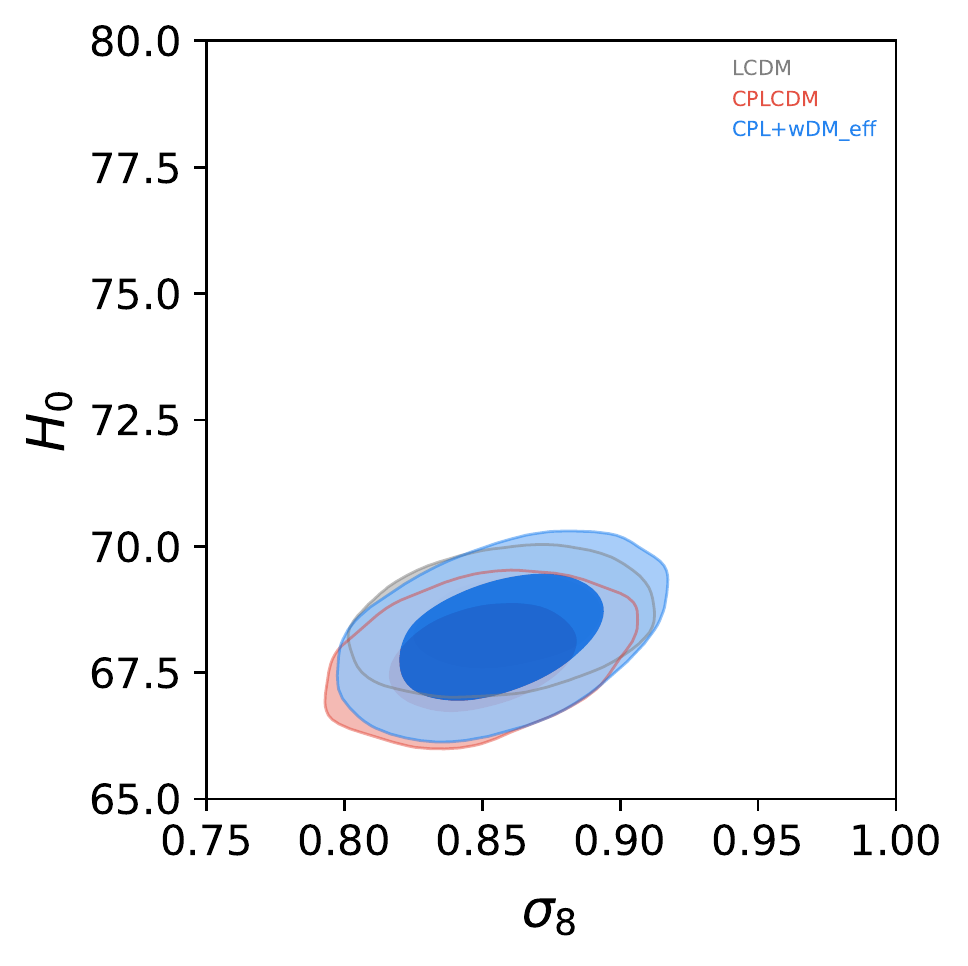} \\
\mbox{(a)} & \mbox{(b)}
\end{array}$
\caption{\footnotesize
$1,2\sigma$ confidence levels in the $H_0-\sigma_8$ parameter spaces
using Planck+R16 (left panel), and Planck+BSH (right panel), for (i)
$\Lambda$CDM (grey), CPLCDM (red) and non-phantom (non)interacting
(blue) models under consideration.}
\label{fig:np_sig_H0}
\end{figure*}

As in the case of phantom EoS, here also a positive correlation
between $H_0$ and $\sigma_8$ is apparent. This has been depicted in
Fig. \ref{fig:np_sig_H0}.  To compare among different datasets, we
have plotted $1,2\sigma$ confidence levels in the $H_0-\sigma_8$
parameter spaces using Planck+R16 in the left panel and Planck+BSH in
the right panel for (i) $\Lambda$CDM (grey), (ii) CPLCDM (red) and
(iii) non-phantom interacting DMDE (blue) models.  These plots reveal a
positive correlation between these two parameters for non-phantom case
as well.

In totality, therefore, we may conclude from the above analysis that
phantom dark energy is preferred over non-phantom by most of the
present datasets except JLA SNe. In the phantom $\wde \leq -1$ regime,
the addition of a very small interaction term ($\wdm \sim -0.001$,
implying transfer of energy from dark energy to dark matter) improves
the fit over $\lam$CDM, and also eases the tension between Planck and
direct $H_0$ measurements, allowing for a very negative
$\wde$. Addition of BAO and SNe causes the equation of state of dark
energy to move closer to $\lam$CDM, thus re-introducing a slight
tension in $H_0$. This is due to inconsistencies within the BSH data:
BAO prefers lower $H_0$ as opposed to R16, SNe does not constrain
$H_0$ but prefers non-phantom DE, and when restricted to phantom and
to $\wdm < 0$ from CMB, increases $\omt$ thereby lowering $H_0$ as
compared to both BAO and R16. The $\sigma_8$ from Planck alone is
lower for phantom models, whereas that for Planck+BSH remains on the
higher end, thus the tension with cluster counts remains for
interacting dark energy models when all data is considered. Overall,
we find that the addition of a small negative $\wdm$, for phantom DMDE
models ($\wde \leq -1$) improves the fit with the data, and eases the
tension between R16 and Planck.  The positive correlation between
$H_0$ and $sigma_8$ appears to be generic to the CMB data, for both
phantom and non-phantom DE EoS. Hence both the tensions cannot be
simultaneously resolved, at least for a wide class of cosmological models using present datasets.


\section{Conclusions}\label{sec:concl}

In this article, we have attempted to investigate the well-known
inconsistencies between different cosmological datasets in a
model-independent framework that takes into account different class of
cosmological models ($\lam$CDM, non-interacting $w_z$CDM, modified
gravity, warm dark matter,  as well as a class of interacting dark matter- dark energy models).  As is well-known,
there is a tension among CMB, R16 and BAO data on preferred values of
$H_0$. Also, CMB data is at odds with cluster data so far as the value
of $\sigma_8$ is concerned. In this article, we tried to check if one
can alleviate these tensions simultaneously, and if so, whether the
choice of cosmological models play a significant role. Our major
findings are summarized below:
\begin{itemize}
\item
A strong positive correlation between $\sigma_8$ and $H_0$ is more or
less generic for the data, irrespective of the choice of cosmological
models ($\Lambda$CDM/ $w_z$CDM/ warm dark matter/a wide class of interacting dark sectors).  The
positive correlation appears to be inbuilt in the CMB data itself, and
is true for both phantom and non-phantom EoS for dark energy.  If one
gets a higher value, the other also shoots up, and vice versa. Since
R16 prefers high $H_0$, and cluster data prefers low $\sigma_8$ as
compared to CMB, both the tensions cannot be simultaneously resolved,
at least using present datasets.
\item
Present data slightly prefers a phantom equation of state for dark
energy and a slightly negative value for effective equation of state for
dark matter (which in turn signifies an energy flow from dark energy
to dark matter and, at the same time, disfavours warm dark matter models,) and for this scenario the use of more parameters opens
up Planck parameter space wide so that high $H_0$ is allowed by
Planck, which is otherwise not achievable in the minimal 6-parameter
$\lam$CDM or CPLCDM cosmology. This comparatively higher value of
$H_0$ is consistent with Planck+R16 data for direct measurement of
$H_0$ but is in tension with BAO and SNe data (and hence with BSH
data) since these prefer a lower value for $H_0$.
\item
Along with a high $H_0$ we also achieve a consistent value for $\omt
\sim 0.3$ for interacting dark sectors.  So, at least, one can resolve
$H_0$ versus Planck tension with a reasonable values for $\omt$ if one
allows interaction, which was not possible to achieve either in
$\lam$CDM model or in non-interacting $w_z$CDM models. This is a clear
advantage of a wide class of interacting DMDE models over the others. 
These models with effective phantom EoS get slight edge over the
others so far as present data are concerned.
\item
Freeing up some parameters (thereby opening up the interacting dark
sector) allows us to have a comparatively lower value of $\sigma_8$
(compared to $\Lambda$CDM or $w_z$CDM) from Planck alone. However,
when Planck data is taken together with BSH, it rises again and become
inconsistent with cluster counts.  A value for $\sigma_8$ which is
consistent with cluster counts is achievable for Planck alone, or with SNe
data, but this would lead to an $H_0$ in tension with both galaxy BAO
and R16 data.
\item  
Thus it is possible to alleviate the tension between the high redshift
CMB data and individual low redshift datasets by changing the expansion
history of the universe to include at least a class of interacting DMDE models. However,
the low redshift data have inconsistencies within themselves so that
it is not possible to match all the low redshift datasets to CMB
simultaneously. Here we have explored these underlying tensions within
low redshift datasets, which have not been explored earlier. For
example, CMB data can get to the high $H_0$ from R16 but this leads to
a high $\sigma_8$ as well, which is problematic from cluster
counts. CMB and SNe data can together achieve low $\sigma_8$ to match
cluster counts, but only for a $H_0$ much lower than that for R16 or
BAO. BAO chooses an $H_0$ that is typically lower than that from R16,
but not low enough to then be consistent with the $\sigma_8$ from cluster
counts. SNe data prefers a non-phantom EoS for DE and is therefore in
tension with most other datasets. So, the usual practice of using BSH
data thereby clubbing R16, BAO and JLA together, with all their
internal inconsistencies, may not be a wise method for the estimation of
cosmological parameters.
\item
For non-phantom (quintessence) case, the results are not too
encouraging, which resonates with earlier findings that a phantom EOS
for dark energy is slightly favoured so far as present data are
concerned. However, for non-phantom case as well, there are direct
indications of a strong positive correlation between $\sigma_8$ and
$H_0$ . This is in tune with our conclusion that one cannot
simultaneously resolve both the tensions, no matter if one considers
$\lam$CDM, non-interacting $w_z$CDM, or a wide class of phantom or non-phantom
interacting dark sectors.
\end{itemize}
In conclusion, we reiterate that: phantom dark energy with energy flow
from dark energy to dark matter is slightly preferred over other
classes of models for the present dataset; the low redshift BSH data
has inconsistencies within itself and with CMB, and using all the data
in conjunction does not necessarily give a true picture of the
universe; and lastly that it is not possible to achieve low $\sigma_8$
and high $H_0$ simultaneously for a wide class of DE models using
present datasets.

\section*{Acknowledgements}

We gratefully acknowledge use of publicly available code COSMOMC. We
also thank computational facilities of ISI Kolkata and IIA Bangalore, 
specially to Mr. Anish Parwage for setting up the computing facilities 
at IIA for COSMOMC runs.
AB thanks DST, India for financial support through INSPIRE fellowship
DST-INSPIRE/IF150497.  UA was purportedly supported through the Young
Scientist Grant YSS/2014/000096 of DST, India.



\begin{thebibliography}{}	

\bibitem[Ade \etal (2016)]{planck15} 
Ade, P. A. R., \etal 2016, \asta, 594, A24.

\bibitem[Aghamousa \etal (2017)]{shaf17_obs} 
Aghamousa, A., Hamann, J., Shafieloo, A. 2017, \jcap, 09, 031 

\bibitem[Alam \etal (2017)]{alam16_sdss12} 
Alam, S., \etal 2017, \mn, 470, 2617 

\bibitem[Alam (2010)]{alam10_obs} 
Alam, U. 2010, \apj, 714, 1460 

\bibitem[Alam \& Lasue (2017)]{alam17_sn} 
Alam, U. \& Lasue, J. 2017, \jcap, 1706, 034

\bibitem[Alam \etal (2017)]{alam17_mg}
Alam, U., Bag, S., \& Sahni, V. 2017, \prd, 9, 023524  

\bibitem[Amendola (1999)]{amen99_dmde} 
Amendola, L. 1999, \prd, 60, 043501 
 
\bibitem[Amendola (2000)]{amen00_dmde}
Amendola, L. 2000 \prd, 62, 043511

\bibitem[Amendola (2004)]{amen04_dmde} 
Amendola, L. 2004, \prd, 69, 103524 

\bibitem[Auborg \etal (2015)]{aub14_bao} 
Aubourg, E., \etal 2015, \prd, 92, 123516

\bibitem[Bahamonde \etal (2017)]{baha17_de} 
Bahamonde, S., \etal 2017, {{\tt arXiv:1712.03107}}.

\bibitem[Bean \etal (2008)]{bean08_dmde} 
Bean, R., Flanagan, E. E., Laszlo, I.  \& Trodden, M. 2008, \prd, 78, 123514 

\bibitem[Betoule \etal (2014)]{bet14_jla} 
Betoule, \etal 2014, \asta, 568, A22 

\bibitem[Beyer \etal (2011)]{wett11_dmde} 
Beyer, J., Nurmi, S. \& Wetterich, C. 2011, \prd,  84, 023010 

\bibitem[Billyard \& Coley (2000)]{coley00_dmde} 
Billyard, A. P. \& Coley, A. A. 2000, \prd, 61, 083503 

\bibitem[Boehmer \etal (2008)]{boeh08_dmde} 
Boehmer, C. G., Caldera-Cabral, G., Lazkoz, R. \& Maartens, R. 2008, \prd, 78, 023505 

\bibitem[Bohringer \etal (2014)]{bohr14_s8} 
Bohringer, H., Chon, G., and Collins, C. A. 2014, \asta, 570, A31

\bibitem[Bonvin \etal (2016)]{bonv16_h0}
Bonvin, V. et al., 2017, \mn, 465, 4914 
 
\bibitem[Busti \& Clarkson (2016)]{clark16_obs} 
Busti, V. C., Clarkson, C. 2016, \jcap, 05, 008 

\bibitem[Chen \etal (2016)]{ratra16_h0} 
Chen, Y., ~Kumar, S. and ~Ratra, B. 2017, \apj, 835, 86 

\bibitem[Chevallier \& Polarski (2001)]{cp01} 
Chevallier, M., \& Polarski, D, 2001, Int. J. Mod. Phys. D, 10, 213 

\bibitem[Chimento \etal (2003)]{pavon03_dmde} 
Chimento, L. P., Jakubi, A. S., Pavon, D. \& Zimdahl W. 2003, \prd, 67, 083513 

\bibitem[Clifton \etal (2012)]{skor12_de} 
Clifton, T., Ferreira, P. G., Padilla, A., \& Skordis, C. 2012, Phys.Rept., 513, 1 

\bibitem[Comelli \etal (2003)]{comel03_dmde} 
Comelli, D., Pietroni, M., \& Riotto, A. 2003, \plb, 571, 115 

\bibitem[Copeland \etal (2006)]{cope06_de}
Copeland, E. J., Sami, M., \& Tsujikawa, S. 2006, Int. J. Mod. Phys. D, 15, 1753 

\bibitem[Damour \& Polyakov (1994)]{dam94_dmde}
Damour, T. \& Polyakov, A. M. 1994, Nucl Phys B, 423, 532

\bibitem[Das \etal (2006)]{das06_dmde} 
Das, S., Corasaniti, P. S., \& Khoury, J. 2006, \prd, 73, 083509 

\bibitem[Delubac \etal (2014)]{del14_baoh} 
Delubac, T. et al. \asta, 574, A59

\bibitem[Di Valentino \etal (2017)]{melc17_obs} 
Di Valentino, E., Melchiorri, A., Linder, E. V., Silk, J. 2017, \prd, 96, 023523 

\bibitem[Di Valentino \etal (2017)]{melc17_dmde} 
Di Valentino, E., Melchiorri, A., \& Mena, O. 2017, \prd, 96, 043503 

\bibitem[Durrer \& Maartens (2010)]{durr08_de} 
Durrer, R. \& Maartens, R. 2010, Dark Energy: Observational \& Theoretical Approaches, ed. P Ruiz-Lapuente (Cambridge UP, 2010), 48 

\bibitem[Efstathiou (2014)]{efst14_h0} 
Efstathiou, G. 2014, \mn, 440, 1138 

\bibitem[Fang \etal (2008)]{hu08_ppf}	
Fang, W., Hu, W. \& Lewis, A. 2008, \prd, 78, 087303

\bibitem[Farrar \& Peebles (2004)]{farr04_dmde} 
Farrar, G. R., \& Peebles, P. J. E. 2004, \apj, 604, 1 

\bibitem[Frieman \etal (2008)]{frie08_de} 
Frieman, J. A., Turner, M. S., \& Huterer, D. 2008, Ann. Rev. \asta, 46, 385 

\bibitem[Gomez-Valent \& Amendola (2018)]{amen18_obs} 
Gómez-Valent, A. and Amendola, L. 2018, \jcap, 1804, 051

\bibitem[Holden \& Wands (2000)]{wands00_dmde} 
Holden, D. J. \& Wands, D. 2000, \prd, 61, 043506 

\bibitem[Holsclaw \etal (2010)]{hols10_obs} 
Holsclaw, T., \etal 2010, \prl, 105, 241302 

\bibitem[Hwang \& Noh (2002)]{hwang02_dmde} 
Hwang, J. C.  \& Noh, H. 2002, Class.Quant.Grav., 19, 527 

\bibitem[Karwal \& Kamionkowski(2016)]{Karwal:2016vyq} 
Karwal, T. \& Kamionkowski, M.\ 2016, \prd, 94, 103523.

\bibitem[Kumar \& Nunes (2017)]{nunes17_dmde} 
Kumar, S. \& Nunes, R. C. 2017, Eur. Phys. J. C,  77, 734 

\bibitem[Lazkoz \etal (2012)]{laz12_obs}
Lazkoz, R., Salzano, V., Sendra, 2012, I., Eur. Phys. J. C, 72, 2130 

\bibitem[Linder (2003)]{lin03} 
Linder, E. V. 2003, \prl, 90, 091301 

\bibitem[Lopez Honorez \etal (2010)]{mena10_dmde} 
Lopez Honorez, L., Mena, O. \& Panotopoulos, G. 2010, \prd, 82, 123525

\bibitem[Mantz \etal (2015)]{mantz15_s8} 
Mantz, A. B. \etal, 2015, \mn, 446, 2205 

\bibitem[Mishra \& Sahni (2018)]{sahni18_dmde} 
Mishra, S. S. \& Sahni, V. 2018, {{\tt arXiv:1803.09767}}.

\bibitem[Moresco \& Marulli (2017)]{mores17_obs} 
Moresco, M., Marulli, F. 2017, \mn, 471, L82 

\bibitem[Mortonson \etal (2014)]{mort14_de} 
Mortonson, M. J., Weinberg, D. H. \& White, M. 2014, {{\tt arXiv:1401.0046}};

\bibitem[Nojiri \& Odinstov (2011)]{noj11_de} 
Nojiri, S. \& Odintsov, S. D. 2011, Phys.Rept., 505, 59 

\bibitem[Padmanabhan (2003)]{paddy03_de} 
Padmanabhan, T. 2003, Phys. Rep.,  380, 235. 

\bibitem[Pavan \etal (2012)]{pavan12_dmde} 
Pavan, A., Ferreira, E. G., Micheletti, S., de Souza, J. \& Abdalla, E. 2012, \prd, 86, 103521 

\bibitem[Peebles \& Ratra (2003)]{peeb03_de} 
Peebles, P. J. E., \& Ratra, B. 2003, Rev. Mod. Phys., 75, 559

\bibitem[Pettorino \etal (2012)]{amen12_dmde}
Pettorino, V., Amendola, L., Baccigalupi, C. \& Quercellini, C. 2012, \prd, 86, 103507 

\bibitem[Pourtsidou \etal (2013)]{skor13_dmde} 
Pourtsidou, A., Skordis,C. \& Copeland, E. 2013, \prd, 88, 083505 

\bibitem[Riess \etal (2011)]{riess11_h0} 
Riess, A. \etal 2011, \apj, 730, 119 

\bibitem[Riess \etal (2016)]{riess16_h0} 
Riess, A.~G. et al., 2016, \apj, 826, 56 

\bibitem[Sahni (2004)]{sahni04_de} 
Sahni, V. 2004, Lect. Notes Phys., 653, 141 

\bibitem[Sahni \& Starobinsky (2000)]{sahni00_de} 
Sahni, V. \& Starobinsky, A. A. 2000, Int. J. Mod. Phys. D, 9, 373. 

\bibitem[Shafieloo \etal (2013)]{shaf13_obs} 
Shafieloo, A., Kim, A. G., Linder, E. V. 2013, \prd, 87, 023520

\bibitem[Tamman \& Reindl (2013)]{tamm13_h0} 
Tammann, G. A. \& Reindl, B. 2013, \asta,  549, A136 

\bibitem[Tarrant \etal (2012)]{cope12_dmde} 
Tarrant, E. R., van de Bruck, C., Copeland, E. J. \& Green, A. M. 2012, \prd, 85, 023503 

\bibitem[Tsujikawa (2010)]{tsuj10_de} 
Tsujikawa, S. 2010, ``Dark Matter and Dark Energy: a Challenge for the 21st Century'', {{\tt arXiv: 1004.1493}};

\bibitem[Valiviita \etal (2008)]{val08_dmde} 
Valiviita, J., Majerotto, E. \& Maartens, R. 2008, \jcap, 0807, 020 

\bibitem[Valiviita \& Palmgren (2015)]{val15_dmde} 
Valiviita, J. \& Palmgren, E. 2015, \jcap, 1507, 015 

\bibitem[Yu \etal (2018)]{ratra18_obs} 
Yu, H., Ratra, B., Wang, F-Y 2018, \apj,  856, 3 

\bibitem[Zhai \etal (2017)]{tink17_obs} 
Zhai, Z., Blanton, M., Slosar, A., Tinker, J. 2017, \apj,  850, 183

\bibitem[Zhao \etal (2012)]{critt12_obs} 
Zhao, G-B, Crittenden, R. G., Pogosian, L., Zhang, X. 2012, \prl, 109, 171301 

\bibitem[Zimdahl \& Pavon (2001)]{pav01_dmde} 
Zimdahl, W. \& Pavon, D. 2001, \plb, 521, 133

\end{thebibliography}
\end{document}